\documentclass{elsart}
\usepackage{hyperref}
\usepackage[square,comma,sort&compress]{natbib}
\usepackage{graphicx}
\usepackage{hyphenat}
\usepackage[latin1]{inputenc}
\usepackage{amssymb}
\newlength{\sepmod}
\newcommand{\ds}{\displaystyle}

\def\l{\langle}

\def\r{\rangle}

\begin{document}

\begin{frontmatter}

\title{The $F$ model on dynamical quadrangulations}
\author[mw]{Martin Weigel}
\ead{weigel@itp.uni-leipzig.de}
\and
\author{Wolfhard Janke}
\ead{janke@itp.uni-leipzig.de}
\thanks[mw]{Present address: Department of Physics, University of Waterloo, Waterloo,
  Ontario, N2L~3G1, Canada}

\address{Institut f\"ur Theoretische Physik, Universit\"at Leipzig, Augustusplatz
  10/11, 04109 Leipzig, Germany}

\begin{abstract}
  The dynamically triangulated random surface (DTRS) approach to Euclidean quantum
  gravity in two dimensions is considered for the case of the elemental building
  blocks being quadrangles instead of the usually used triangles. The well-known
  algorithmic tools for treating dynamical triangulations in a Monte Carlo simulation
  are adapted to the problem of these {\em dynamical quadrangulations\/}. The thus
  defined ensemble of 4-valent graphs is appropriate for coupling to it the 6- and
  8-vertex models of statistical mechanics.  Using a series of extensive Monte Carlo
  simulations and accompanying finite-size scaling analyses, we investigate the
  critical behaviour of the 6-vertex $F$ model coupled to the ensemble of dynamical
  quadrangulations and determine the matter related as well as the graph related
  critical exponents of the model.
\end{abstract}


\end{frontmatter}

\section{Introduction}
\label{sec:intro}

Einstein gravity being perturbatively non-renormalizable as a field theory,
constructive approaches towards a quantization of gravity have been an ever more
active field of research in the past decades \citep{smolin:03a}. The dynamical
triangulations model in its Euclidean and Lorentzian versions has proved a successful
ansatz for the formulation of such a consistent theory of quantum gravity
\citep{ambjorn:book,ambjorn:00c}. Compared to the more fancy methods, such as string
theory \citep{witten:99a} and non-commutative geometry \citep{connes:00a}, it is
rather more minimalistic in trying to directly model the quantum fluctuations of
space-time by a probabilistic sum over an ensemble of discrete, simplicial manifolds
\citep{weingartenundco}. For the Euclidean case in two dimensions, this ensemble can
be defined as the set of all gluings of equilateral triangles to a regular, usually
closed surface of fixed topology, while counting each of the possible gluings with
equal weight. The resulting random-surface model and its simplicial generalisation to
higher dimensions are numerically tractable, for instance by Monte Carlo simulations.
Furthermore, for the case of two dimensions the use of matrix models and
generating-function techniques led to exact solutions for the cases of pure Euclidean
gravity \citep{brezin:78a,boulatov:86a} and the coupling of certain kinds of matter,
such as the Ising model \citep{kazakov:86a,boulatov:87a,burda}, to the surfaces.
These two-dimensional theories generically exhibit continuous phase transitions on
tuning the relevant coupling parameters accordingly and thus allow for taking the
intended continuum limit. In the case of matter variables coupled to two-dimensional
dynamical triangulations, the critical exponents governing the transitions are
conjectured exactly from conformal field theory as functions of the exponents on
regular lattices via the so-called KPZ/DDK formula \citep{kpzddk}
\begin{equation}
  \label{eq:kpz}
  \tilde \Delta = {\sqrt{1 - C + 24 \Delta} - \sqrt{1-C} \over \sqrt{25 -C} -
    \sqrt{1 -C}},
\end{equation}
where $\Delta$ is the original scaling weight, $\tilde \Delta$ the scaling weight
after coupling to gravity and $C$ the central charge. The field-theory ansatz leading
to Eq.\ (\ref{eq:kpz}) breaks down for central charges $C>1$, an effect which has
been termed the $C=1$ ``barrier'', whereas the discrete model of $C>1$ matter coupled
to dynamical triangulations stays well defined. This mismatch of descriptions and its
driving mechanism is still one of the rather poorly understood aspects of the
dynamical triangulations model \citep{david:97a,ambjorn:00c,ambjorn:00d}.

Ice-type or vertex models on {\em regular\/} lattices form one of the most general
classes of models of statistical mechanics with discrete symmetry (for reviews see,
e.g., Refs.\ \citep{lieb:domb,baxter:book}). Special cases of this class of models
can be mapped onto more well-known problems such as Ising and Potts models or graph
colouring problems \citep{baxter:book}. For the case of two-dimensional lattices,
several of these vertex models can be solved exactly, yielding a very rich and
interesting phase diagram including various transition lines as well as critical and
multi-critical points \citep{baxter:book}. Thus, for two-dimensional vertex models
one has the rare combination of a rich structure of phases and an exceptional
completeness of the available analytical results. Hence, coupling this class of
models to a fluctuating geometry of the dynamical triangulations type is of obvious
interest, both as a prototypic model of statistical mechanics subject to annealed
connectivity disorder and as a paradigmatic type of matter coupled to two-dimensional
Euclidean quantum gravity. Recently, the use of matrix model methods led to a
solution of the thermodynamic limit of a special 6-vertex model, the $F$ model,
coupled to planar $\phi^4$ graphs \citep{zinn}. It was found to correspond to a $C=1$
conformal field theory, i.e., it lies on the boundary to the region $C>1$, where the
KPZ/DDK solution \citep{kpzddk} breaks down. Also, a special slice of the 8-vertex
model could be analysed via transformation to a matrix model \citep{kazakov:99a}. A
generalisation of this result to the general parameter space of the 8-vertex model is
currently being attempted \citep{ambjorn:01a,zinn-justin:03a}.  However, owed to the
method of matrix integrals, these studies neither reveal the behaviour of the matter
related observables and the details of the occurring phase transitions nor the
fractal properties of the graphs such as, e.g., their Hausdorff dimension.
Especially, for the case of the 6-vertex model, which turns out to exhibit a phase
transition of the Berezinskii-Kosterlitz-Thouless (BKT) type, quantities related to
the staggered, anti-ferroelectric order parameter cannot be easily constructed, such
that a detailed numerical analysis of the problem seems valuable.  Numerically it is
found here that, due to the combined effect of the presence of logarithmic
corrections to scaling expected for a $C=1$ theory and the comparative smallness of
the effective linear extent of the accessible graph sizes, the leading scaling
behaviour is obscured by extremely strong finite-size corrections. Thus, a very
careful scaling analysis incorporating the various correction terms has to be
performed in order to disentangle the corrections from the asymptotic scaling form.

Since the 6- and 8-vertex models of statistical mechanics are defined on a lattice
with four-valent vertices, instead of considering dynamical triangulations or the
dual planar, ``fat'' (i.e., orientable) $\phi^3$ graphs, one has to use an ensemble
of dynamical {\em quadrangulations\/} or the dual $\phi^4$ Feynman diagrams as the
geometry to model the coupling of vertex models to quantum gravity.  This can be
rather easily done within the framework of matrix model methods
\citep{brezin:78a,thooft:99a}. For Monte Carlo studies, however, it turns out that
the well established simulation techniques for dynamical triangulations
\citep{boulatov:86a,pachner:91a,ambjorn:94a} are quite cumbersome to adapt to the
case of four-valent graphs which, therefore, only very scarcely have been considered
in the literature \citep{johnston:93a,baillie:95a}. Especially, ergodicity for the
selected set of moves has to be ensured and a method of coping with the observed
severe critical slowing down of the dynamics, such as an adaption of the
``baby-universe surgery'' method \citep{ambjorn:94a,ambjorn:95e}, has to be devised.
The details of these modifications to the simulation scheme will be presented in a
separate publication \citep{prep}.

The rest of this paper is organised as follows. In Sec.\ \ref{sec:vertex} we first
review the basic properties of vertex models on regular lattices. We shortly discuss
the matrix-model solution of the 6-vertex model and elaborate on the necessary
conceptual and simulational modifications for considering vertex models on random
graphs. Section \ref{sec:pt} is devoted to an in-depth investigation of the BKT phase
transition of the 6-vertex $F$ model coupled to planar, ``fat'' $\phi^4$ graphs by
means of an extensive series of Monte Carlo simulations.  In Sec.\ \ref{sec:geom} we
present our numerical results for the geometrical properties of the coupled system,
such as the string susceptibility exponent and the internal Hausdorff dimension.
Finally, Sec.\ \ref{sec:concl} contains our conclusions.

\section{Vertex models on random graphs}
\label{sec:vertex}

\subsection{Vertex models on regular lattices}
\label{sec:vertreg}

An {\em ice-type\/} or {\em vertex\/} model was first proposed by Pauling
\citep{pauling} as a model for (type I) ice. In this model, the two possible
positions of the hydrogen atoms on the bonds of the crystal formed by the oxygens, if
symbolised by arrows, lead to six different allowed configurations around a vertex
provided that the experimentally observed {\em ice rule\/} is satisfied, stating that
each vertex has two incoming and two outgoing arrows, see, e.g., Ref.\
\citep{baxter:book}.  While for the original ice model all vertex configurations were
counted with equal probability, for the general 6-vertex model vertex energies
$\epsilon_i$ are introduced, resulting in Boltzmann factors
$\omega_i=\exp(-\epsilon_i/k_BT)$, where $T$ denotes temperature and $k_B$ is the
Boltzmann constant. Some symmetry relations are commonly assumed between the weights
$\omega_i$; in particular, given the interpretation of the arrows as electrical
dipoles, in the absence of an external electric field the partition function should
be invariant under a simultaneous reversal of all arrows, leading to the identities
$a=\omega_1=\omega_2$, $b=\omega_3=\omega_4$ and $c=\omega_5=\omega_6$.  An
especially symmetric version of the model assumes $\epsilon_a=\epsilon_b=1$,
$\epsilon_c=0$ or $a=b$, $c=1$. This so-called $F$ model \citep{rys:63a} turns out to
exhibit an {\em anti-\/}ferroelectrically ordered ground state. The square-lattice,
zero-field 6-vertex model has been solved exactly in the thermodynamic limit by means
of a transfer matrix technique (Bethe ansatz) by Lieb \citep{lieb} and Sutherland
\citep{sutherland:67a}. The analytic structure of the free energy is most
conveniently parameterised in terms of the variable \citep{baxter:book}
\begin{equation}
  \Delta=\frac{a^2+b^2-c^2}{2ab}. 
  \label{eq:delta_def}
\end{equation}
The free energy takes a different analytic form depending on whether $\Delta<-1$,
$-1<\Delta<1$ or $\Delta>1$. Thus, phase transitions occur, whenever $|\Delta|=1$.
The case $\Delta>1$ corresponds to two symmetry-related ferroelectrically ordered
phases termed I and II, $\Delta<-1$ denotes an anti-ferroelectrically ordered phase
IV and $-1<\Delta<1$ is attained in the disordered phase III. The latter phase has
the peculiarity of having an infinite correlation length throughout, which can be
traced back to the fact that it corresponds to a critical surface of the more general
8-vertex model \citep{baxter:book}. From Eq.\ (\ref{eq:delta_def}) it is obvious that
the $F$ model exhibits a phase transition on cooling down from the
infinite-temperature point $a=b=c=1$ contained in the disordered phase III to
somewhere in the anti-ferroelectrically ordered phase IV. The transitions I
$\rightarrow$ III and II $\rightarrow$ III are first-order phase transitions
\citep{baxter:book}. The transition III $\rightarrow$ IV of the $F$ model, on the
other hand, exhibits an essential singularity of the free energy known as the BKT
phase transition \citep{kt}.

While the ferroelectrically ordered phases exhibit an overall polarisation which can
be used as an order parameter for the corresponding transition, the
anti-ferro\-electric order of phase IV is accompanied by a {\em staggered\/}
polarisation with respect to a sub-lattice decomposition of the square lattice. That
is, when decomposing the square lattice into two new square lattices tilted by
$\pi/4$ against the original one, the anti-ferroelectric ground states correspond to
a {\em ferroelectric\/} ordering of the vertices of the sub-lattices with opposite
signs of the overall polarisation of the sub-lattices. An order parameter for the
corresponding transition can be defined by introducing overlap variables $\sigma_i$
for each vertex of the lattice such that $\sigma_i = v_i\ast v_i^0$, where $v_i$
denotes the arrow configuration at vertex $i$, $v_i^0$ one of the two
anti-ferroelectric ground-state configurations and the product ``$\ast$'' symbolises
the overlap given by
\begin{equation}
  v\ast v' \equiv \sum_{k=1}^4 A_k(v) A_k(v'),
\end{equation}
where $k$ numbers the four edges around each vertex and $A_k(v)$ should be $+1$ or
$-1$ depending on whether the corresponding arrow of $v$ points out of the vertex or
into it \citep{baxter:book}. Then, the {\em spontaneous staggered polarisation\/}
$P_0=\l\sigma_i\r/2=\l\sigma\r/2$ vanishes in the disordered phase and approaches
unity in the thermodynamic limit for low temperatures in phase IV and can thus be
used as an order parameter for the anti-ferroelectric transition. 

Vertex models on regular lattices are closely linked with different series of
integrable models, which in turn are related to an exhaustive enumeration of certain
conformal field theories. In fact, it turns out that the 6-vertex model, being the
critical version of the 8-vertex model, includes in suitable generalisations the
critical points of all of the well-known two-dimensional lattice models of
statistical mechanics, including the Ising and Potts models as the most prominent
examples. Especially, the restricted solid-on-solid (RSOS) models
\citep{andrews:84a}, which realise each central charge of the unitary series of
minimal models \citep{huse:84a}, have been shown to asymptotically map onto the
8-vertex model, such that the critical RSOS models correspond to 6-vertex models.
Furthermore, an impressive series of models in two dimensions can be mapped onto the
Coulomb gas \citep{nienhuis:domb}. In these mappings, an intermediate step is always
given by models of the solid-on-solid (SOS) type, which again can be related to
vertex models \citep{beijeren:77a}. Combining these methods, the 6-vertex model can
be described as the common element among critical systems in two dimensions
\citep{pasquier}.

\subsection{Vertex models on random lattices}
\label{sec:vertrand}

Putting a vertex model onto a {\em random\/} four-valent graph such as the quantum
gravity $\phi^4$ graphs imposes an additional restriction on the class of vertex
weights that can be sensibly considered. The ferroelectrically ordered phases I and
II of the 6-vertex model and the order parameter describing the corresponding phase
transition depend on the existence of a global notion of direction. On a random
graph, this notion is maldefined. The only local orientational structure available is
that of the vertices and faces of the graph. Thus, for an 8-vertex model coupled to
quantum-gravity $\phi^4$ random graphs, one has to assume that $a=b$, while the other
vertex types can still be distinguished with only a cyclic ordering of the links
around each vertex. For the 6-vertex model this leaves only two fundamentally
different choices of models to be sensibly considered: the $F$ model with
$\epsilon_a=\epsilon_b=1$, $\epsilon_c=0$ and the so-called {\em inverse F (IF)
  model\/} with $\epsilon_a=\epsilon_b=-1$, $\epsilon_c=0$, which, however, is not of
much interest here due to its lack of an ordered phase.

\begin{figure}[tb]
  \centering
  \includegraphics[clip=true,keepaspectratio=true,width=8cm]{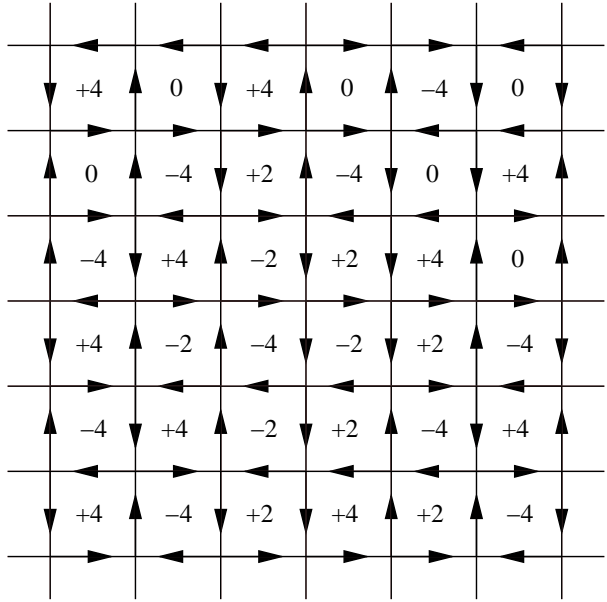}
  \caption
  {Transformation of the square-lattice 6-vertex model to a ``spin'' model on the
    dual lattice. The four links of each plaquette of the lattice are traversed
    counter-clockwise. The ``spin'' values written in the centres of the plaquettes
    are oriented sums of $\pm 1$ around the plaquettes. Thus, the occurring ``spin''
    values are $0,\pm2,\pm4$.}
  \label{fig:plaquette}
  \vspace*{0.15cm}
\end{figure}

For the square lattice an order parameter for the anti-ferroelectric transition of
the $F$ model could be defined by a suitably calculated overlap between the actual
state and one of the two anti-ferroelectrically ordered ground states. On a random
graph, the corresponding ground states are not so easily found and, moreover, vary
between different realisations of the connectivity of the graph.  Hence, to define an
anti-ferroelectric order parameter for the random graph case, a different and more
suitable representation of the vertex model has to be sought.  Above, the
anti-ferroelectrically ordered state has been described as mutually opposite
ferroelectric order on two complementary sub-lattices. A decomposition of the square
lattice of this kind corresponds to a {\em bipartition\/} or {\em two-colouring\/} of
its sites. Unfortunately, the considered random $\phi^4$ graphs are not bipartite in
general, preventing an immediate application of this prescription. When interpreting
the vertex-model arrows as a discrete vector field on the lattice, the ice rule for
the 6-vertex model translates to a zero-divergence condition for this field. We thus
transform the vertex model from its interpretation as a field on the links of the
original lattice to a representation of the curl of this field on the faces of the
lattice or, equivalently, the sites of the dual lattice. Following Stokes' theorem,
this is done by integrating the vertex model arrows around the elementary plaquettes.
By convention, plaquettes are traversed counter-clockwise, adding $+1$ for each arrow
pointing in the direction of motion and $-1$ otherwise. On the square lattice the
resulting ``spins'' (or ``heights'') on the plaquettes can assume the values $0$,
$\pm2$, $\pm4$.  This is demonstrated in Fig.\ \ref{fig:plaquette}. In this way, the
6-vertex model can be transformed to a sort of ``spin model'' on the dual of the
original lattice.  Note, however, that one has rather involved restrictions for the
``spin'' values allowed between neighbouring plaquettes, which would lead to quite
cumbersome interaction terms when trying to write down a Hamiltonian.

In the new representation, the anti-ferroelectrically ordered state of the model
again has a sub-lattice structure. However, in contrast to the sub-lattice
decomposition of the original representation, now the {\em dual\/} lattice is broken
down into ``black'' and ``white'' sub-lattices, such that no two plaquettes of the
same colour share a link. Then, an order parameter for the anti-ferroelectric
transition can be defined as the thermal average of the sum of the plaquette
``spins'', e.g., on the ``black'' plaquettes. Reflecting the construction of the
plaquette ``spins'' in Fig.\ \ref{fig:plaquette} it is obvious that this definition
of the order parameter {\em exactly\/} coincides with the original definition of
Sec.\ \ref{sec:vertreg} on the level of configurations.  The difference is, however,
that the new definition can be easily generalised to the case of arbitrary lattices,
as long as their {\em duals\/} are bipartite. This is the case for the planar random
$\phi^4$ graphs we are considering since any planar quadrangulation is bipartite.
Thus, we can introduce a two-colouring of the faces of the graphs. While for the
square lattice the numbers of black and white plaquettes are always the same, the
black and white faces of the $\phi^4$ random graphs not necessarily occur at equal
proportions. Thus, one should take the ``spins'' of both types of faces into account,
however ``weighted'' with the colour of the faces.  Therefore, the configurational
value of the staggered polarisation of the $F$ model on a planar $\phi^4$ random
graph ${\mathcal G}$ can be defined as $P \equiv \frac{1}{2}\sum_{v\in V({\mathcal
    G}^\ast)}C_v S_v$, where ${\mathcal G}^\ast$ denotes the dual of the graph,
i.e.\, the quadrangulation, $V({\mathcal G}^\ast)$ the set of vertices of ${\mathcal
  G}^\ast$, $C_v=\pm1$ the ``colour'' of the plaquette of ${\mathcal G}$
corresponding to the vertex $v$ of ${\mathcal G}^\ast$ and $S_v$ the plaquette
``spin'' at $v$. Recalling the construction of the plaquette ``spins'', this can also
be written in terms of the $\phi^4$ graph ${\mathcal G}$ as
\begin{equation}
   P = \frac{1}{2}\sum_{f\in F({\mathcal G})}\sum_{l_f\in f}C_f A(l_f),
  \label{eq:disorderd_op_2}
\end{equation}
where $F({\mathcal G})$ denotes the set of faces of ${\mathcal G}$, $l_f$ the links
of face $f$, $C_f=\pm 1$ the ``colour'' of $f$ and $A(l_f)=\pm 1$ the direction of
the vertex-model arrow on link $l_f$ with respect to the prescribed anti-clockwise
traversal of the faces. The thermal average $\l P\r/2$ is now taken as the order
parameter of a possibly occurring anti-ferroelectric phase transition of the $F$
model coupled to planar $\phi^4$ random graphs. Note, however, that due to the
overall arrow reversal symmetry of the vertex model the expectation value $\l P\r$
will vanish at any temperature for a finite graph. Thus, for finite graphs we
consider the modulus $\l|P|\r$ instead, analogous to the usual treatment of the
magnetisation of the Ising model.

As mentioned above in the Introduction, a matrix model related to the $F$ model
coupled to planar $\phi^4$ random graphs could be solved exactly in the thermodynamic
limit \citep{zinn}. The solution is related to a transformation of the $F$ model to a
model of close-packed loops by using ``breakups'' of the vertices, i.e.,
prescriptions for connecting incoming and outgoing arrows. The original weights of
the 6-vertex model translate into weights for the oriented loops by assigning a phase
factor $\exp( i\mu\pi/2)$ to each left turn and a phase factor $\exp(-i\mu\pi/2)$ to
each right turn of an oriented loop \citep{baxter:76a,baxter:book}. Here, the
coupling $\mu$ is related to the weights of the $F$ model as\footnote{Note that, in
  terms of the parameter $\Delta$ of Eq.\ (\ref{eq:delta_def}), this choice of
  weights covers only the range $-1<\Delta<1$, which corresponds to the disordered
  phase of the square-lattice $F$ model.}
\begin{equation}
  a/c=b/c=[2\cos(\pi\mu)]^{-1}. 
  \label{eq:sixvertex_lambda}
\end{equation}
On the square lattice the phase factors around each loop always multiply up to a
total of $\exp(\pm i\mu2\pi)$ due to the absence of curvature.  On a random graph,
however, a loop $l$ in general receives a non-trivial weight $\exp[i\mu\Gamma(l)]$
with $\Gamma(l)$ denoting the integral of the geodesic curvature along the curve $l$,
i.e.,
\begin{equation}
  \Gamma(l) = \frac{\pi}{2}\left(\mbox{\# left turns}-\mbox{\# right turns}\right). 
\end{equation}
This loop expansion is related to the well-known loop representation of the O($n$)
model of Ref.\ \citep{domany:81a}. There, on a regular lattice, due to the absence of
curvature all loops receive the same constant fugacity $n=2\exp(\pm i\mu2\pi)$,
leading to the critical O($n$) model. On the considered random graphs this picture
only remains valid for the limiting case $\mu=0$, where the curvature dependence
cancels. Thus, the $\mu=0$ point of the $F$ model on random planar $\phi^4$ graphs is
equivalent to the critical O($2$) loop model \citep{kostov:89b,dalley:92a,zinn} and
thus, by universality, the critical $XY$ model\footnote{Note that the loops occurring
  in the expansion of the O($n$) model are not in general close packed on the lattice
  as are the loops of the presented loop expansion of the $F$ model. However, the
  critical O($2$) model lies at the boundary of the dense phase of the O($n$) model,
  where loops are close packed \citep{kostov:92b}.}. Note that this corresponds to
the same critical point $a/c=b/c=1/2$ as on the regular square lattice, which is
natural since the symmetry breaking is induced by the choice of the vertex weights.
By means of the mentioned matrix model techniques it is found that the $F$ model
coupled to planar, ``fat'' $\phi^4$ graphs has a critical point for each value of the
coupling $\mu$ (corresponding to the disordered phase III), in agreement with the
behaviour on the square lattice. Exploring the vicinity of this critical point, it is
found that the string susceptibility exponent $\gamma_s=0$ for all $\mu$, leading to
only logarithmic divergences of the free energy \citep{zinn}. This behaviour is
indeed expected from the $C\rightarrow 1$ limit of the KPZ/DDK prediction Eq.\
(\ref{eq:kpz}). Thus, the general phase structure of the $F$ model coupled to planar
random $\phi^4$ graphs in the grand-canonical ensemble of a varying number of
vertices has been found in Ref.\ \citep{zinn}. The existence of a BKT type phase
transition at $\mu=0$ was obvious beforehand from the equivalence to the O(2) loop
model at this point. Details of the behaviour of matter-related observables close to
the critical point, such as the scaling of the staggered anti-ferroelectric
polarisability, however, could naturally not be extracted from the matrix model
ansatz.

\section{The anti-ferroelectric phase transition}
\label{sec:pt}

The critical point of the $F$ model on the square lattice provided the first example
of an infinite-order phase transition of the BKT type. By virtue of the loop
expansion sketched above, this behaviour is expected to persist as the model is
coupled to a random lattice. In the vicinity of a phase transition of this type, the
usual thermal and finite-size scaling (FSS) relations are profoundly changed. Using
an elaborate set of simulational techniques specially tailored for simulations of
this model, we present a detailed scaling analysis of its thermal properties.  As a
guideline for the rather involved analysis we used our newly performed set of
loop-cluster update simulations of the square-lattice $F$ model \citep{weigel:05a},
which is computationally less demanding such that much larger system sizes could be
investigated. For a general discussion of scaling and FSS at an infinite-order phase
transition of the BKT type \citep{kt}, we refer the reader to this study and
references found therein \citep{weigel:05a}. Due to the nature of the occurring
singularities the main strengths of FSS are found not to apply to the BKT phase
transition, and the focus of numerical analyses of the $XY$ and related models has
been on {\em thermal\/} scaling, see, e.g., Ref.\ \citep{gupta}. In addition,
renormalization group analyses predict {\em logarithmic corrections\/} to the leading
scaling behaviour \citep{amit}, as expected for a $C=1$ theory, which have been found
exceptionally hard to reproduce numerically due to the presence of higher order
corrections of comparable magnitude \citep{wj:97b}.  Comparing the phase transitions
in the two-dimensional planar and the six-vertex $F$ models, one should keep in mind
that due to the dual relation of both models, the r\^oles of high- and
low-temperature phases are exchanged in that the $F$ model has a critical
low-temperature phase, whereas the high-temperature phase is massless in the $XY$
model. In contrast to the $XY$ model, the low-temperature phase of the $F$ model
exhibits a non-vanishing order parameter, given by the spontaneous polarisation of
Eq.\ (\ref{eq:disorderd_op_2}), such that, although the critical points of both
models are equivalent, the magnetisation of the planar model does not correspond to
the polarisation of the $F$ model \citep{weigel:05a}.

\subsection{Simulation techniques}
\label{sec:technique}

For Monte Carlo simulations of two-dimensional combinatorial dynamical triangulations
or the dual regular $\phi^3$ graphs, an ergodic set of updates for simulations of a
fixed number of polygons or graph vertices (canonical ensemble) is given by the
so-called Pachner moves \citep{gross:92a}. An adaption of the link-flip move for
canonical simulations of triangulations to the case of quadrangulations has been
proposed in Refs.\ \citep{johnston:93a,baillie:95a}. By the construction of
counter-examples it can be shown that the link-flip moves of Refs.\
\citep{johnston:93a,baillie:95a} do {\em not\/} in general constitute an ergodic
dynamics for canonical simulations of dynamical quadrangulations. Introducing a
second type of link-flip moves, we construct an algorithm for canonical simulations
of dynamical quadrangulations, which does not show any signs of ergodicity breaking
\citep{weigel:02b,doktor,prep}. A scaling analysis of the thus constructed dynamics
reveals that its performance --- as expected from a local algorithm --- is limited by
the effect of critical slowing down. To alleviate this problem, we adapt the
non-local ``baby-universe surgery'' method proposed in Ref.\ \citep{ambjorn:94a} for
triangulations to the case of quadrangulations and investigate its dynamical
properties by means of a scaling analysis \citep{doktor,prep}. For the vertex model
part, we also employ a non-local, cluster algorithm known as ``loop-cluster
algorithm'', which is known to drastically reduce autocorrelation times for vertex
models on the square lattice \citep{evertz:loop}. For the application of this
simulation scheme to random-lattice models, certain modifications are necessary. The
mentioned algorithmic developments for the graph and the vertex model part as well as
the technical details of the necessary simulational set-up will be discussed in a
separate publication \citep{prep}.

For the FSS study to be presented below, we simulated a series of spherical $\phi^4$
graphs of sizes ranging from $N_2=256$ up to $N_2=65\,536$ vertices\footnote{The use
  of the variable $N_2$ for this number has its origin in the general notation for
  simplicial manifolds, where $N_d$ denotes the number of $d$-simplices of the
  simplicial complex, see, e.g., Ref. \citep{ambjorn:book}.}. The simulations where
performed at several computing facilities using about 100\,000 hours of CPU time in
total.

\subsection{Scaling analysis}
\label{sec:results1}

We assume a parameterisation of the $F$ model coupling parameters which involves a
temperature variable and thus sticks more closely to the language of statistical
mechanics than to that of field theory. It hence differs from the parameterisation
(\ref{eq:sixvertex_lambda}) used in the context of the matrix model solution, which
only covers the critical disordered phase of the $F$ model. Choosing the vertex
energies as $\epsilon_a=\epsilon_b=1$, we have $a=b=e^{-\beta}$, $c=1$, where
$\beta=1/k_BT$, such that the BKT point occurs for $\beta_c=\ln 2$, both for the
square-lattice model and, conjectured by the matrix model solution discussed in Sec.\
\ref{sec:vertrand}, for the $F$ model coupled to planar $\phi^4$ random graphs.

\subsubsection{The specific heat}

The specific heat $C_v$ of the $F$ model coupled to planar $\phi^4$ random graphs
exhibits a broad peak of around $C_v\approx 0.45$, shifted away from the critical
point into the low\hyp{}temperature phase\footnote{Note that the specific heat of the
  2D $XY$ model exhibits a peak in the {\em high-temperature\/} phase, as expected
  from duality.} to a centre of $\beta\approx 1.0$. The peak does not depend on the
lattice size up to very small finite-size corrections, i.e., no FSS is observed. The
expected essential, non-divergent singularity \citep{lieb:domb,weigel:05a} cannot in
general be resolved, since it is covered by the presence of non-singular background
terms.  This non-scaling behaviour of the specific heat is commonly considered as a
first good indicator for the presence of an infinite-order phase transition
\citep{barber:domb}.

\subsubsection{Location of the critical point}

The critical coupling can be determined from the scaling of the shifts of suitably
defined pseudo-critical couplings on finite graphs, see Ref.\ \citep{weigel:05a}.
Here, we use the locations $\beta_\chi$ of the maxima of the staggered
anti-ferroelectric polarisability, defined from the generalised polarisation of Eq.\
(\ref{eq:disorderd_op_2}). In terms of the inverse temperature $\beta$ to first order
one has at a BKT transition \citep{weigel:05a},
\begin{equation}
  \beta_\chi(N_2) = \beta_c + A_\beta(\ln N_2)^{-1/\rho},
  \label{rand_temp_shifts_simple}
\end{equation}
where $N_2$ is the size of the graphs and $\rho = 1/2$ for the regular $XY$ and $F$
models \cite{barber:domb,baxter:book}. For the determination of the peak positions we
make use of the temperature-reweighting technique \citep{ferrenberg}. Note that the
quoted errors do not cover the potential bias induced by the reweighting procedure.
We performed simulations for graph sizes between $N_2=256$ and $N_2=25\,000$ sites,
taking some $10^6$ measurements after the systems had been equilibrated. Measurements
were taken after every tenth sweep of the combined link-flip and ``baby-universe
surgery'' dynamics, using ``regular'' graphs without self-energy and tadpole
insertions \citep{prep}. All statistical errors were determined by a combined
binning/jackknife technique, cf.\ Ref.~\citep{efron}.

\enlargethispage{0.4cm}

Comparing the estimated peak locations to the corresponding results for the
square-lattice model \citep{weigel:05a} one notes that the accessible part of the
scaling regime is strongly shifted towards lower temperatures, being rather far away
from the conjectured critical coupling $\beta_c=\ln 2 \approx 0.693$, cf.\ the
``regular ensemble'' data of Fig.\ \ref{phi4_peaks3} below. We start with fits of the
simple form Eq.~(\ref{rand_temp_shifts_simple}) without including any correction
terms.  Additionally, we assume $\rho=1/2$ here as in the square-lattice case, which
has to be justified {\em a posteriori\/} by the thermal scaling analysis. Within this
scheme, the influence of correction terms is taken into account by successively
omitting lattice sizes from the small-$N_2$ side. Due to the strong corrections
present, however, no fits with satisfactory fit quality can be found in this way such
that it appears mandatory to include correction terms. Since the exact form of the
present scaling corrections is not known, an effective description has to be
employed. One possible ansatz is to relax the constraint $\rho=1/2$, introducing
$\tilde{\rho}\neq\rho$ as an additional fit parameter. Even for this type of fit,
acceptable fit qualities can only be attained by dropping many of the smaller graph
sizes, thus strongly increasing the uncertainty in the estimated parameters.
Additionally, we find that the fit results for small minimum included graph sizes
$N_{2,\mathrm{min}}$ partly depend on the choice of the starting values for the fit
parameters, i.e., that the fit routine gets stuck in local minima of the $\chi^2$
distribution. For $N_{2,\mathrm{min}} = 4096$ we arrive at an estimate $\beta_c =
0.83(58)$, $A_\beta=1.7(62)$, and $1/\tilde{\rho}=1.0(31)$ with a quality of
$Q=0.69$. Statistically, this is in agreement with the expected value $\beta_c=\ln
2\approx 0.693$ for the critical coupling, but due to the large statistical error the
estimate is of limited significance. The result for the exponent $\tilde{\rho}$
cannot be taken as a serious estimate for $\rho$, since it incorporates corrections
effectively.

\begin{table}[tb]
  \centering
  \caption
  {Parameter results of linear fits of the form
    (\ref{rand_temp_shifts_simple3}) to the simulation data for the peak locations of the
    staggered polarisability. Values of parameters held fixed are indicated by square
    brackets.} 
  \vspace*{0.25cm}
    \begin{tabular}{|r|r@{.}l|r@{.}l|r@{.}l|r|r@{.}l|} \hline
      \multicolumn{1}{|c|}{$N_{2,\mathrm{min}}$} & \multicolumn{2}{|c|}{$\beta_c$} &
      \multicolumn{2}{|c|}{$A_\beta$} & \multicolumn{2}{|c|}{$B_\beta$}&
      \multicolumn{1}{|c|}{$C_\beta$} & \multicolumn{2}{|c|}{$Q$} \\ \hline \hline
      256  &  0&8999(71) &  17&4(11)   &  $-69$&4(47)    &   \multicolumn{1}{|c|}{[0]}   &  0&02  \\
      512  &  0&876(13)  &  21&9(22)   &  $-92$&0(108)   &   \multicolumn{1}{|c|}{[0]}   &  0&08  \\
     1024  &  0&817(24)  &  33&7(46)   & $-155$&3(243)   &   \multicolumn{1}{|c|}{[0]}   &  0&72  \\ \hline
      256  &  0&779(39)  &  55&4(122)  & $-424$&9(1140)  & 918(294) & 0&32 \\
      512  &  0&693(75)  &  87&0(263)  & $-748$&1(2647)  & 1838(741) & 0&39 \\ \hline 
    \end{tabular}
  \vspace*{0.15cm}
  \label{phi4_F_maxima_simple2_table}
\end{table}

For the square-lattice case, from the exact solution the leading corrections to the
form (\ref{rand_temp_shifts_simple}) with $\rho=1/2$ could be expressed as a power
series in $1/\ln N_2$ \cite{weigel:05a},
\begin{equation}
  \beta_\chi(N_2) = \beta_c + A_\beta(\ln N_2)^{-2} + B_\beta(\ln N_2)^{-3}
  + C_\beta(\ln N_2)^{-4},
  \label{rand_temp_shifts_simple3}  
\end{equation}
hence we consider this form for the random graph data here as well. As can be seen
from the collection of fit parameters in Table \ref{phi4_F_maxima_simple2_table},
this form provides a good description of the data, although some of the statistical
errors of the fit parameters become very large. Neglecting the second correction
first, i.e., holding $C_\beta = 0$ fixed, the results are stable on successively
omitting data points from the small-$N_2$ side, and the resulting estimates for the
transition temperature are slowly drifting towards the asymptotic value $\beta_c =
\ln 2 = 0.693\ldots$. Nevertheless, the result for, e.g., $N_{2,\mathrm{min}}=1024$,
$\beta_c=0.817(24)$ is still far from being compatible with the asymptotic result in
terms of the statistical error. Including the fourth-order term of
(\ref{rand_temp_shifts_simple3}), on the other hand, further reduces the estimates
for $\beta_c$ to the extent of being compatible with $\beta_c=\ln 2$, however at the
price of largely increased statistical errors. For $N_{2,\mathrm{min}} > 512$, the
fits get very unstable, such that we quote as our final result from this approach
$\beta_c=0.693(75)$ for $N_{2,\mathrm{min}} = 512$. If we finally {\em fix\/}
$\beta_c$ at its asymptotic value, for $C_\beta=0$ we reach a fit quality of $Q=0.01$
only at $N_{2,\mathrm{min}} = 2048$, while with variable $C_\beta$, $Q=0.52$ is
reached already at $N_{2,\mathrm{min}} = 512$. This clearly shows that {\em both\/}
correction terms are necessary for resolving the scaling corrections, but the
accuracy of the present data is only marginally sufficient to do so. It should be
noted that also the other types of fits presented here still yield good
quality-of-fits when fixing the parameter $\beta_c$ at $\ln 2$. For example, a fit of
the form (\ref{rand_temp_shifts_simple}) with variable exponent $\tilde{\rho}$ to the
data with $N_{2,\mathrm{min}}=2048$ gives $A_\beta = 1.071(81)$, $1/\tilde{\rho} =
0.541(35)$, and $Q = 0.84$.

\subsubsection{Universality of the critical coupling\label{sec:universality}}

One might be tempted to suspect that the observed rather large deviations of the
finite-size positions of the polarisability maxima from the expected value
$\beta_c=\ln 2\approx0.693$ are due to the fact that we use graphs of the {\em
  regular\/} ensemble, i.e.\ those without self-energy and tadpole insertions,
whereas the matrix model calculations of Ref.\ \citep{zinn} naturally concern graphs
of the unrestricted {\em singular\/} ensemble. Indeed, quite generally one does {\em
  not\/} expect the critical coupling of a model to be {\em universal\/}. In
particular, for the Ising model coupled to dynamical polygonifications or the dual
graphs, the location of the observed transition does depend on whether one considers
spins located on the vertices of triangulations, quadrangulations, $\phi^3$ or
$\phi^4$ graphs \citep{boulatov:87a,johnston:93a,baillie:95a}. Additionally,
depending on the considered ensemble of graphs with respect to the inclusion or
exclusion of certain types of singular contributions, one arrives at different values
for the critical coupling \citep{boulatov:87a,jurkiewicz:88a,burda,schneider:99a}.
However, the situation is quite different for the case of the $F$ model coupled to
random lattices. As has been mentioned above in Sec.\ \ref{sec:vertrand}, in the
matrix model description of the problem, the matrix potential becomes equivalent to
that of the O($2$) model in the limit $\mu=0$ \citep{zinn}, which corresponds to the
choice $a/c=b/c=1/2$ or $\beta_c=\ln 2$. Thus, renormalizing the matrix model to
remove some or all of the singular graph contributions does not change the location
of the BKT point.

\begin{figure}[tb]
  \centering
  \includegraphics[clip=true,keepaspectratio=true,width=12cm]{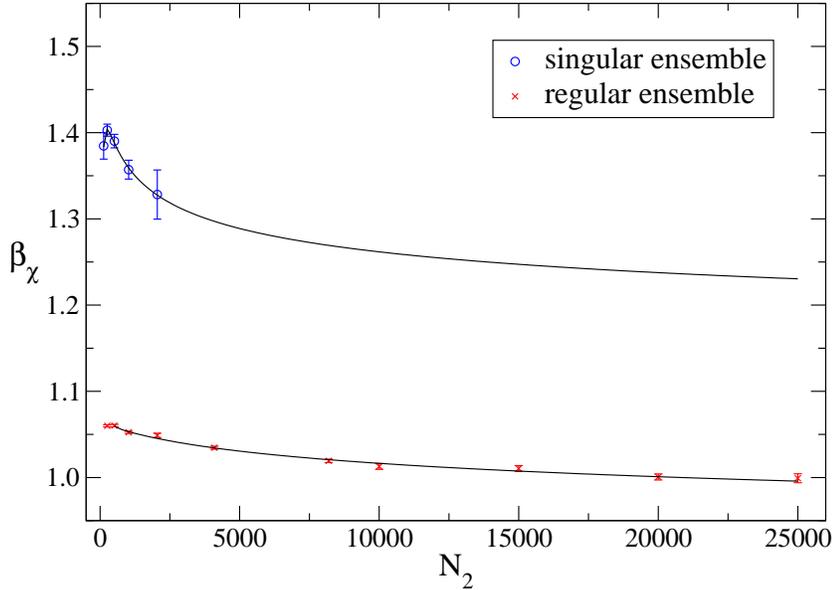}
  \caption
  {Finite-size approach of the peak locations of the staggered polarisability of the
    $F$ model on $\phi^4$ random graphs with (``singular ensemble'') and without
    (``regular ensemble'') tadpole and self-energy insertions. The solid lines show
    fits of the functional form (\ref{rand_temp_shifts_simple3}) to the data.} 
  \label{phi4_peaks3}
  \vspace*{0.15cm}
\end{figure}

We have not performed extensive simulations of graphs of the ``singular'' ensemble
including self-energy and tadpole insertions to demonstrate this behaviour
numerically. This is due to the fact that simulations for graphs of the singular
ensemble are by orders of magnitude less efficient for the considered graph sizes
than simulations of the other graph ensembles due to details of the implementation of
the simulation scheme, cf.\ Ref.\ \citep{prep}. Nevertheless, we carried out some
simulations for smaller graph sizes and analysed the FSS of the peak locations of the
staggered polarisability just as for the case of ``regular'' graphs. The
corresponding FSS data are shown in Fig.\ \ref{phi4_peaks3} together with the results
for regular graphs. Using Eq.~(\ref{rand_temp_shifts_simple3}) with $C_\beta=0$, a
fit to the data including all five points from $N_2=128$ to $N_2=2048$ yields the
estimate $\beta_c = 1.01(11)$, $Q = 0.92$, letting $C_\beta$ vary gives $\beta_c =
0.83(69)$, $Q = 0.76$, which is in principle in agreement with $\beta_c=\ln 2$,
although very inaccurate. Note that from Fig.\ \ref{phi4_peaks3} the finite-size
corrections for the singular graph case are much larger than those for the regular
graph model.  This is in contrast to previous observations for the case of the Potts
model coupled to random triangulations \citep{ambjorn:95a} and the resulting common
belief that the inclusion of singular graph contributions generically reduces FSS
corrections.

\begin{figure}[tb]
  \centering
  \includegraphics[clip=true,keepaspectratio=true,width=12.5cm]{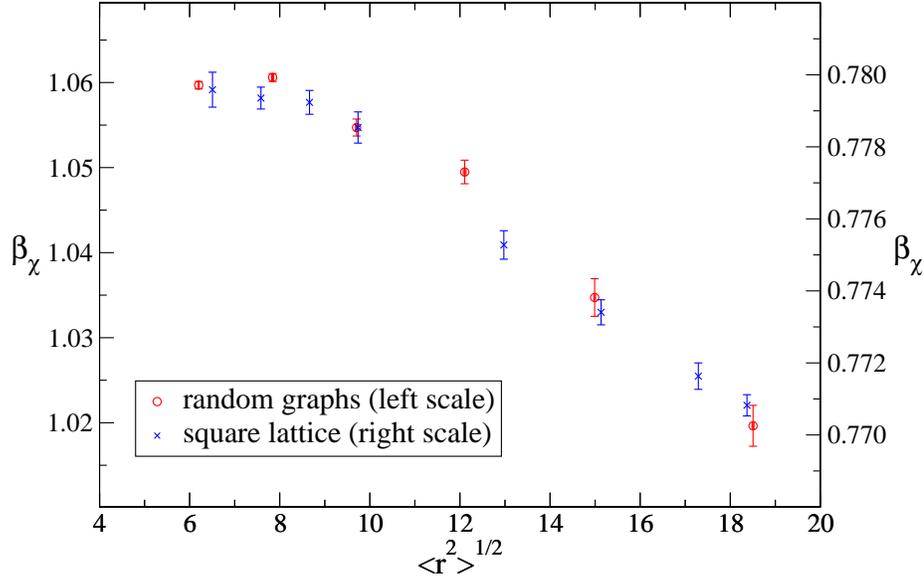}
  \caption
  {Collapse of the FSS approach of the scaling of the peak
    locations of the staggered anti-ferroelectric polarisability of the $F$ model on
    random $\phi^4$ graphs (left scale) and on the square lattice (right scale). The
    data for the square-lattice model are taken from a set of simulations presented
    in Ref.\ \citep{weigel:05a}.} 
  \label{peaks_scaling_comparison}
  \vspace*{0.15cm}
\end{figure}

As has been mentioned in the Introduction, the reason for the observed very slow
approach to the expected asymptotic behaviour lies in the double effect of the
presence of logarithmic corrections to scaling and the small effective linear extent
of the highly fractal lattices. In principle it should be possible to resolve the
resulting scaling corrections by including higher-order correction terms in the fit
ans\"atze.  However, it must be admitted that, refraining from any artificial
``good-will'' tinkering with the fit parameters, the accuracy of the present data is
not sufficient for reliable many-parameter, possibly non-linear fits. The strength of
this combined effect is nicely demonstrated numerically by the fact that the fits to
the FSS of the polarisability peak locations with $\beta_c$ fixed to its true value
$\beta_c=\ln 2$ come as close as $\beta_\chi(N_2)=0.7$ to the critical value only for
graph sizes $N_2\approx 10^{100}$ for the form (\ref{rand_temp_shifts_simple3}) with
variable $C_\beta$ or even $N_2\approx 10^{5000}$ for the form
(\ref{rand_temp_shifts_simple}) with variable exponent $\tilde{\rho}$. Instead of
figuring out more elaborate fits, we try to disentangle the two correction effects by
a comparison to the square-lattice model, where only the logarithmic corrections are
present, but the considered lattices are not fractal \citep{weigel:05a}. For this
purpose, we plot in Fig.~\ref{peaks_scaling_comparison} the polarisability peak
locations as a function of the root mean square extent of the considered lattices
defined as
\begin{equation}
  \l r^2\r_{N_2}^{1/2} = \left\l\frac{\sum_{r=0}^{r_\mathrm{max}}r^2 G_{11}(r)}
    {\sum_{r=0}^{r_\mathrm{max}}G_{11}(r)}\right\r_{N_2}^{1/2},
  \label{msqe_def_sim}
\end{equation}
which is the relevant measure for the linear extent of the graphs. Here, we take the
geometrical two-point function $G_{11}(r)$ as the number of graph vertices with a
geodesic link distance $r$ from a randomly chosen reference point $p_0$. The root
mean square extents $\l r^2\r^{1/2}$ are related to the number of graph vertices
according to $\l r^2\r \sim N_2^{2/d_h}$, which defines the internal Hausdorff
dimension $d_h$. Due to the fractal structure of the random graphs, largely differing
values of the root mean square extent $\l r^2\r^{1/2}$ are found for them in
comparison to square lattices with the same number of vertices $N_2$. For the latter
data (which are basically exact) this scaling ansatz without inclusion of any
correction terms yields $d_h=2.000(20)$, where the error reflects discretisation
effects for small lattices.  For the case of $\phi^4$ random graphs the fit yields
$d_h=3.336(11)$.  Note, however, that the result for $d_h$ is slowly increasing as
more and more of the small-$N_2$ lattices are excluded and we expect the true value
of the Hausdorff dimension to be somewhat larger, see Refs.\
\citep{kawamoto:92a,ambjorn:97b,ambjorn:98d} and Sec.\ \ref{sec:geom} below. Hence,
in order to obtain results for the $F$ model at comparable linear extents of the
square and random lattices, one has to consider rather small volumes for the
square-lattice case. For the comparison we use $L\times L$ square lattices, where the
edge lengths $L$ were chosen such that the resulting root mean square extent comes as
close as possible to the $\l r^2\r^{1/2}$ values for the corresponding $\phi^4$
random graphs. The volumes of the $\phi^4$ random graphs were chosen between
$N_2=256$ and $N_2=8192$, increasing in powers of two.

\enlargethispage{0.5cm}

In Fig.\ \ref{peaks_scaling_comparison} we present a comparison of the FSS approach
of the peak locations of the polarisability for the $\phi^4$ graph and square-lattice
\citep{weigel:05a} models plotted as a function of the linear extent $\l r^2\r^{1/2}$
of the lattices.  Here, the abscissae of the plot have been scaled such as to account
for the difference in the overall correction amplitude, but {\em assuming the same
  value $\ln 2$ for the offset\/}. From the two simulation points near ${\l
  r^2\r}^{1/2}\approx 10$ we find the ratio of the correction amplitudes
as\footnote{These two simulation points have been chosen since there the difference
  in $\l r^2\r^{1/2}$ between the square and random lattices is minimal within the
  set of considered lattice sizes.}
\begin{equation}
  A_\beta = \frac{\beta_\chi^\mathrm{rl}(N_2=1024)-\ln 2}
  {\beta_\chi^\mathrm{sl}(N_2=324)-\ln 2} \approx 4.23,
\end{equation}
where $\beta_\chi^\mathrm{rl}$ denotes the peak position for the random $\phi^4$
graph model and $\beta_\chi^\mathrm{sl}$ the value for the square lattice. The thus
achieved collapse of the FSS data is obvious from Fig.\
\ref{peaks_scaling_comparison}. Consequently, we come to the clear conclusion that
the larger deviations of the peak locations for random graphs are simply due to an
about four times larger overall amplitude of the correction terms as compared to the
square-lattice model, the details of the FSS approach being otherwise surprisingly
similar between the two considered lattice types. Especially, the fact that for the
$\phi^4$ graph case the asymptotic value $\beta_c=\ln 2$ cannot be clearly resolved
by the considered fits to the data is an obvious consequence of the comparative
smallness of the accessible lattice sizes in terms of their effective linear extents
$\l r^2\r^{1/2}$. To underline this finding, we performed fits of the simple form
(\ref{rand_temp_shifts_simple}) to the data for both types of lattices (there are not
enough data points for fits with correction terms), including sizes starting from the
points near $\l r^2\r^{1/2}\approx 10$, which result in estimates
$\beta_c=0.7554(18)$ for the square lattice and $\beta_c=0.9416(89)$ for the random
graphs.  In terms of the quoted statistical errors these are obviously both far away
from the asymptotic result. The deviation from $\beta_c=\ln 2$ is, however, just
about four times larger for the random graph case than for the square-lattice model,
in agreement with the previous discussion of the scaling collapse of Fig.\
\ref{peaks_scaling_comparison}.

\subsubsection{Critical energy and specific heat}

\begin{figure}[tb]
  \centering
  \includegraphics[clip=true,keepaspectratio=true,width=12cm]{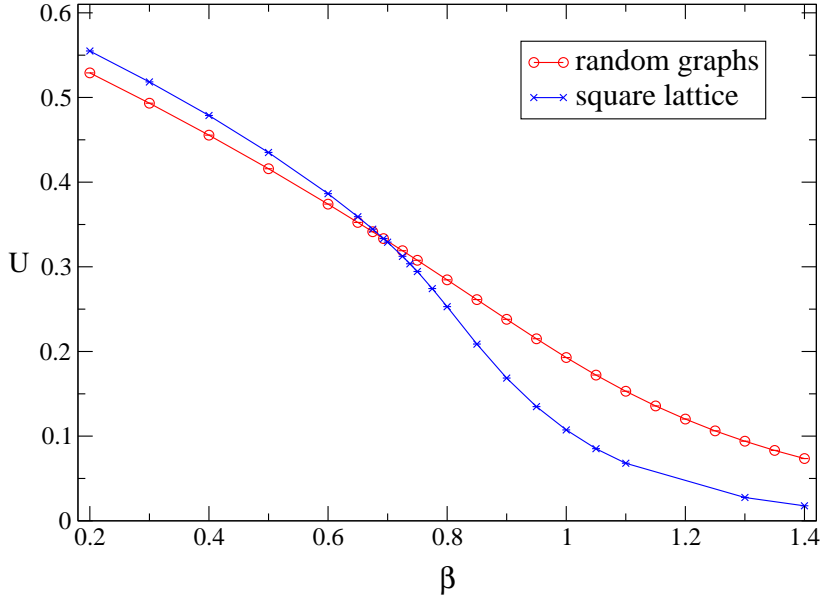}
  \caption
  {Temperature dependence of the internal energy $U$ of the square-lattice and random
    $\phi^4$ graph $F$ models. Simulations have been performed for a $N_2=46^2=2116$
    square lattice and random graphs with $N_2=2048$ sites.} 
  \label{phi4_energy_comp}
  \vspace*{0.15cm}
\end{figure}

As an aside, we note that for the largest $\phi^4$ random graphs we have simulated,
i.e., for $N_2=65\,536$, at $\beta=\beta_c=\ln 2$ we find the following values of the
internal energy and specific heat per site,
\begin{equation}
    U(\beta=\ln 2) = 0.333355(11),\;\;\; C_v(\beta=\ln 2) = 0.2137(12). 
\end{equation}
Comparing these results to the values found analytically for the square-lattice $F$
model \citep{lieb:domb}, $U(\beta_c) = 1/3$, $C_v(\beta_c) = 28(\ln
2)^2/45\approx0.2989$, we see that $U(\beta=\ln 2)$ is very close to the value found
for the square lattice, whereas $C_v(\beta=\ln 2)$ is far away from the
square-lattice result. On the basis of these findings, we conjecture that the
critical value of the internal energy of the $F$ model is not affected by the
coupling to random graphs, while the critical specific heat is. Thus, as one would
expect, the critical distribution of vertex energies naturally changes its shape on
moving from the square-lattice to the random graph model, but, curiously, its mean is
not shifted by this procedure.  Interestingly, this situation seems to be specific to
the critical point $\beta_c=\ln 2$ common to both models, where the two curves cross. For
other inverse temperatures the square-lattice and random graph energies diverge, see
Fig.\ \ref{phi4_energy_comp}. This probably indicates the presence of an additional
symmetry at criticality.

\subsubsection{FSS of the polarisability}

On coupling the vertex model to quantum gravity we expect a renormalization of the
critical exponents as prescribed by the KPZ/DDK formula (\ref{eq:kpz}). In
Ref.~\citep{kpzddk} KPZ/DDK focus on conformal minimal models with $C<1$ coupled to
the Liouville field, but their work should also marginally apply to the limiting case
$C=1$ of the model considered here. The KPZ/DDK formula prescribes a dressing of the
conformal weights on coupling a matter system to the fluctuating background. To find
the usual critical exponents from the weights, one assumes that the well-known
scaling relations stay valid and thus arrives at
\begin{equation}
  \begin{array}{l}
    \ds\alpha = \frac{1-2\Delta_\epsilon}{1-\Delta_\epsilon},\;\;\;
    \beta = \frac{\Delta_{P}}{1-\Delta_\epsilon},\;\;\;
    \gamma = \frac{1-2\Delta_{P}}{1-\Delta_\epsilon},\\
    \ds d_h\nu = \frac{1}{1-\Delta_\epsilon},\;\;\;
    2-\eta = (1-2\Delta_{P})d_h. 
  \end{array}
  \label{exps_from_weights}
\end{equation}
Here, $\Delta_\epsilon$ denotes the weight of the energy operator and $\Delta_{P}$
symbolises the weight of the scaling operator corresponding to the spontaneous
staggered polarisation $P_0$, which here takes on the r\^ole of the magnetisation
operator $\sigma$ of magnetic models. For the special case of the infinite-order
phase transition considered here, the usual exponents written above are not
well-defined in the sense of describing power-law singularities. However, the
corresponding FSS exponents, i.e., $\beta/d_h\nu = \Delta_{P}$ and $\gamma/d_h\nu =
1-2\Delta_{P}$, still have a well-defined meaning. From the exponent
$\beta/d_h\nu=1/4$ for the square-lattice $F$ model (with $d_h=d=2$)
\citep{weigel:05a}, we find $\Delta_{P} = 1/4$. Note that this weight is different
from the weight $\Delta_\sigma=1/16$ found for the magnetisation of the critical $XY$
model in two dimensions, see e.g.\ Ref.\ \citep{henkel:book}. For central charge
$C=1$ from Eq.\ (\ref{eq:kpz}) one arrives at $\tilde{\Delta}_{P}=1/2$ and the
dressed critical exponents become $\beta/d_h\nu = \tilde{\Delta}_{P} = 1/2$ and
$\gamma/d_h\nu = 1-2\tilde{\Delta}_{P} = 0$, implying a merely logarithmic
singularity of the staggered polarisability for dynamical graphs.

For a numerical check of these conjectured exponents, there are the two principal
possibilities of considering the FSS of the staggered polarisability at its maxima
for the finite graphs {\em or\/} at the fixed asymptotic transition coupling
$\beta_c=\ln 2$. While in the asymptotic regime both approaches are expected to lead
to identical results, this is not at all obvious in the presence of large, not
completely controlled correction effects for the accessible graph sizes. In both
cases, by analogy to the situation on the square lattice \citep{weigel:05a} we start
from an FSS form including a leading effective correction term, namely,
\begin{equation}
   \chi(N_2) = A_\chi N_2^{\gamma/d_h\nu}(\ln N_2)^{\omega_\chi},
  \label{pol_random_simple_fit}
\end{equation}
where $\chi(N_2)$ is taken to be either the peak value as a function of $\beta$ or
the value at $\beta=\beta_c=\ln 2$. We consider the peak value case first, taking the
simulation results for the graph sizes $N_2=256,\ldots,25\,000$.  Omitting the
correction term, i.e., forcing $\omega_\chi=0$, and trying to control the effect of
corrections to scaling by successively omitting data points from the small-$N_2$
side, results in quite poor fits with an exponent estimate $\gamma/d_h\nu\approx 0.7$
steadily decreasing with increasing lower cut-off $N_{2,\mathrm{min}}$. Allowing the
effective correction exponent $\omega_\chi$ to vary, the resulting leading exponent
estimate $\gamma/d_h\nu$ is considerably reduced, still showing a tendency to decline
as $N_{2,\mathrm{min}}$ is increased, cf.\ Table
\ref{phi4_pol_maxima_simple_table}(a). However, the fit quality is still not very
good and the resulting exponent estimate for, e.g., $N_{2,\mathrm{min}}=2048$,
$\gamma/d_h\nu=0.301(79)$ is not consistent in terms of the statistical error with
the purely logarithmic singularity expected from the KPZ/DDK prediction. These
results in principle might be improved by including corrections of the form $1/(\ln
N_2)^n$, $n=1,2,\ldots$ as in the square-lattice case \cite{weigel:05a}, but the
present data are not precise enough to reliably fit these terms.

\begin{table}[tb]
  \centering
  \caption
  {Results of fits of the functional form (\ref{pol_random_simple_fit}) to the
    simulation data for the staggered polarisability. (a) Fits to the data at the
    polarisability peak locations. (b) Fits to the data at the asymptotic
    critical coupling $\beta=\beta_c=\ln 2$.} 
  \vspace*{0.25cm}
  \begin{tabular}{rc}
    \raisebox{8.75ex}{(a)} &
    \begin{tabular}{|r|r@{.}l|r@{.}l|r@{.}l|r@{.}l|} \hline
      \multicolumn{1}{|c|}{$N_{2,\mathrm{min}}$} & \multicolumn{2}{|c|}{$A_\chi$} &
      \multicolumn{2}{|c|}{$\gamma/d_h\nu$} & \multicolumn{2}{|c|}{$\omega_\chi$} &
      \multicolumn{2}{|c|}{$Q$} \\ \hline \hline
      256 & 0&1975(97) & 0&4749(81) & 1&698(55) & 0&00 \\
      512 & 0&116(14) & 0&406(16) & 2&22(12) & 0&00 \\
      1024 & 0&039(12) & 0&281(37) & 3&24(30) & 0&24 \\
      2048 & 0&047(37) & 0&301(79) & 3&07(68) & 0&16 \\ \hline
    \end{tabular}    
    \\ & \\
    \raisebox{7ex}{(b)} &    
    \begin{tabular}{|r|r@{.}l|r@{.}l|r@{.}l|r@{.}l|} \hline
      \multicolumn{1}{|c|}{$N_{2,\mathrm{min}}$} & \multicolumn{2}{|c|}{$A_\chi$} &
      \multicolumn{2}{|c|}{$\gamma/d_h\nu$} & \multicolumn{2}{|c|}{$\omega_\chi$} 
      & \multicolumn{2}{|c|}{$Q$} \\ \hline \hline
      256 & 0&491(19) & 0&0194(55) & 2&117(40) & 0&66 \\
      512 & 0&543(42) & 0&0304(91) & 2&026(72) & 0&91 \\
      1024 & 0&569(75) & 0&035(14) & 1&98(12) & 0&85 \\ \hline
    \end{tabular}
  \end{tabular}
  \vspace*{0.15cm}
  \label{phi4_pol_maxima_simple_table}
\end{table}

For the data at fixed coupling $\beta_c=\ln 2$, simulations up to slightly larger
graph sizes could be performed since no reweighting analysis is necessary there.
Hence, results are available for graph sizes between $N_2=256$ and $N_2=32\,768$
sites, increasing by powers of two. For the constrained fits of the functional form
(\ref{pol_random_simple_fit}) with $\omega_\chi=0$ we do not find a quality-of-fit of
at least $10^{-2}$ for $N_{2,\mathrm{min}}$ up to $4096$ and thus do not consider
this form further. The parameters of fits including the logarithmic term are
collected in Table \ref{phi4_pol_maxima_simple_table}(b), revealing that the
functional form including a logarithmic correction fits the data rather well already
for quite small values of $N_{2,\mathrm{min}}$, leading to exponent estimates
$\gamma/d_h\nu$ compatible with the conjecture $\gamma/d_h\nu=0$ in terms of the
quoted statistical errors. In fact, if we {\em assume\/} a purely logarithmic
increase of $\chi(N_2)$, i.e., if we fix $\gamma/d_h\nu=0$, the data yield
good-quality fits for $N_{2,\mathrm{min}}\gtrsim 512$; for $N_{2,\mathrm{min}}=2048$
the parameters of this purely logarithmic fit are $A_\chi = 0.3960(96)$, $\omega_\chi
= 2.295(11)$, with $Q = 0.39$.

\begin{figure}[tb]
  \centering
  \includegraphics[clip=true,keepaspectratio=true,width=12cm]{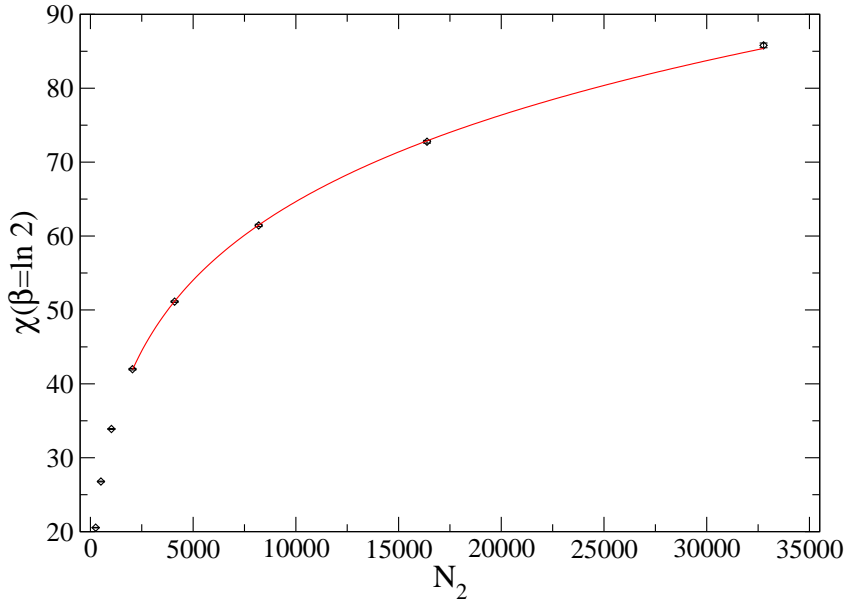}
  \caption
  {Finite-size simulation data of the polarisability of the $F$ model on random
    $\phi^4$ graphs at the asymptotic critical coupling $\beta_c=\ln 2$. The solid
    curve shows a fit of the form (\ref{pol_random_simple_fit}) to the data, where
    $\gamma/d_h\nu=0$ was kept fixed.} 
  \label{phi4_pol_peaks_fig2}
  \vspace*{0.15cm}
\end{figure}

The simulation data at $\beta=\ln 2$ together with this last fit are shown in Fig.\
\ref{phi4_pol_peaks_fig2}. Note that for the peak-height data discussed before, such
a purely logarithmic fit is {\em not\/} possible with acceptable values of $Q$. To
enable a somewhat better judgement of the observed discrepancy between the scaling at
the peak maxima and at $\beta_c=\ln 2$, we considered the same two lines for the {\em
  square-lattice\/} model \citep{weigel:05a}, using a range of lattice sizes
comparable to that of the random graph case in terms of the effective linear extents
as it has been discussed in Sec.~\ref{sec:universality}. Fitting the functional form
(\ref{pol_random_simple_fit}) with variable $\omega_\chi$ to these two square-lattice
data sets, we find $\gamma/d_h\nu=0.475(46)$ for the scaling at $\beta=\ln 2$ also
considered above, but an estimate of $\gamma/d_h\nu=0.598(36)$ from the scaling of
the peak values of $\chi$. Thus, also for the square-lattice model, the scaling of
the peak values yields an exponent estimate lying off the expected result
($\gamma/d_h\nu=1/2$ in this case), while fits at the critical coupling are in good
agreement with the expectations. This is in agreement with the general observation of
enhanced correction amplitudes of the random graph model compared to the
square-lattice case reported in Sec.~\ref{sec:universality}.

\subsubsection{FSS of the spontaneous polarisation}

For the scaling of the spontaneous polarisation the situation is found to be quite
similar to the above discussed case of the polarisability. We assume the same leading
FSS form as in the square-lattice case \citep{weigel:05a}, i.e.,
\begin{equation}
  P_0(N_2) = A_{P_0}N_2^{-\beta/d_h\nu}(\ln N_2)^{\omega_{P_0}},
  \label{phi4_magn_fit_form}
\end{equation}
where, again, $P_0(N_2)$ is taken to be either the value at the peak position of the
polarisability or, alternatively, the result at the asymptotic critical coupling
$\beta_c=\ln 2$. Fits without the logarithmic correction term ($\omega_{P_0}=0$) show
unacceptable quality throughout the whole region of choices of the cut-off
$N_{2,\mathrm{min}}$ and for both FSS series. For the polarisation at the peak
locations of the polarisability, even fits including the logarithmic correction term
of Eq.\ (\ref{phi4_magn_fit_form}) show very poor fit quality and estimates for
$\beta/d_h\nu$ which are clearly too small compared to the KPZ/DDK prediction
$\beta/d_h\nu=1/2$ in terms of their statistical errors. We attribute this to the
generally more pronounced corrections for the values at the polarisability peak
locations already noted above. In addition, however, the non-divergent behaviour of
the polarisation makes it even harder to resolve the correction terms properly, and
the possible presence of systematic reweighting errors (bias) has much more severe
effects here due to the higher statistical accuracy of the polarisation estimate.
Again, the analogous analysis of the FSS of the square-lattice model reveals a
similar behaviour for comparable graph sizes in terms of the linear extent, however
with the size of the deviations from the expected result being much smaller.

\begin{table}[tb]
  \centering
  \caption
  {Parameters resulting from fits of the form (\ref{phi4_magn_fit_form}) to the
    finite-graph spontaneous polarisation at the infinite-volume critical coupling
    $\beta_c=\ln 2$.} 
  \vspace*{0.25cm}
  \begin{tabular}{|r|r@{.}l|r@{.}l|r@{.}l|r@{.}l|} \hline
    \multicolumn{1}{|c|}{$N_{2,\mathrm{min}}$} & \multicolumn{2}{|c|}{$A_{P_0}$} &
    \multicolumn{2}{|c|}{$\beta/d_h\nu$} & \multicolumn{2}{|c|}{$\omega_{P_0}$} &
    \multicolumn{2}{|c|}{$Q$} \\ \hline \hline
    256 & 1&583(35) & 0&4633(30) & 0&726(22) & 0&74 \\
    512 & 1&658(68) & 0&4581(50) & 0&684(39) & 0&91 \\
    1024 & 1&58(11) & 0&4633(79) & 0&728(64) & 0&98 \\
    2048 & 1&48(23) & 0&469(15) & 0&779(134) & 1&00 \\ \hline
  \end{tabular}    
  \vspace*{0.15cm}
  \label{phi4_magn_maxima_simple_table}
\end{table}

Table \ref{phi4_magn_maxima_simple_table} shows the parameters resulting from
least-squares fits of Eq.~(\ref{phi4_magn_fit_form}) to the simulation data at the
fixed coupling $\beta=\beta_c=\ln 2$. The overall quality of the fits is much better
than for the data at the polarisability peak locations discussed before. This is at
least partially due to the fact that for the results at fixed coupling no bias
effects induced by a reweighting procedure are present. We do not observe a clear
overall drift of the exponent estimate $\beta/d_h\nu$ resulting from the fits as a
function of the cut-off $N_{2,\mathrm{min}}$ and the quality-of-fit is found to be
exceptionally high already for small values of $N_{2,\mathrm{min}}$. The result for
$N_{2,\mathrm{min}}=2048$ is consistent with the KPZ/DDK conjecture
$\beta/d_h\nu=1/2$ within about two times the quoted standard deviation. We note that
the estimated correction exponents $\omega_\chi$ and $\omega_{p_0}$ are found to
be clearly different from each other. In fact, from the exact solution of the
square-lattice model, both exponents are found to be different even asymptotically
\citep{weigel:05a}. In addition, both exponents effectively capture the presence of
sub-leading corrections for the two observables, leading to the occurrence of further
differences.

\subsubsection{Thermal scaling}

In order to extract information about the critical exponent $\rho$ and possibly to
find additional evidence for the location of the critical point, we tried to perform
a thermal scaling analysis and considered the dependence of the staggered
anti-ferroelectric polarisability on the inverse temperature $\beta$ in the vicinity
of the critical point. Since the high-temperature phase of the $F$ model coupled to
$\phi^4$ random graphs is expected to be critical as for the case of the
square-lattice $F$ model, such a scaling analysis has to be performed on the
low-temperature side of the polarisability peak. As for the square-lattice model
\citep{weigel:05a}, we find scaling throughout the high-temperature phase. Due to an
exponential slowing down of the link-flip and ``baby-universe surgery'' dynamics of
the $\phi^4$ graphs above $\beta_c$ \citep{prep}, simulations cannot proceed
arbitrarily deep into the ordered phase. Up to the attainable inverse temperatures of
about $\beta=1.4$, we still observe strong finite-size effects and no asymptotic
collapse of the curves for different graph sizes, which is again attributed to the
large fractal dimension of the graphs.

\begin{figure}[tb]  
  \centering
  \includegraphics[clip=true,keepaspectratio=true,width=12cm]{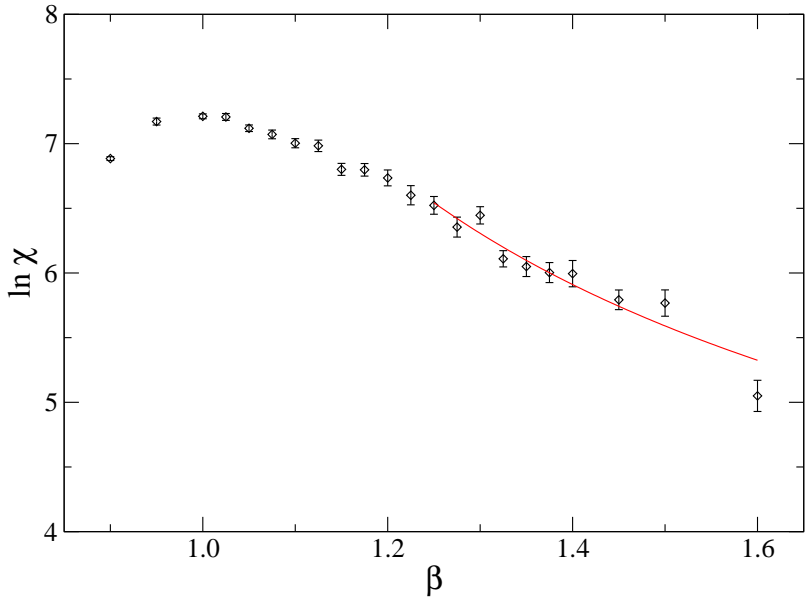}
  \caption
  {Thermal scaling of the polarisability of the random graph $F$ model for graphs
    with $N_2=30\,000$ sites. The curve shows a fit of the function
    (\ref{thermal_scaling_phi4_formula}) to the data, where $\beta_c=\ln 2$ and
    $\rho=1/2$ have been kept fixed.} 
  \label{phi4_thermal_scaling_fit}
  \vspace*{0.15cm}
\end{figure}

The requirements of a proper thermal scaling analysis of the polarisability resulting
from these observations are almost impossible to fulfil: one has to keep enough
distance from the critical point for the linear extent of the graph to be large
compared to the correlation length of the matter part to keep finite-size effects
under control and, on the other hand, one should not proceed too deep into the
ordered phase such as not to leave the thermal scaling region in the vicinity of the
critical point. Thus, one would have to go to huge graph sizes to get rid of these
constraints to a practically acceptable extent. Nevertheless, we attempt a thermal
scaling analysis of the polarisability from simulations of graphs of size
$N_2=30\,000$ with inverse temperatures ranging from $\beta=0.9$ up to $\beta=1.6$
taking about $800\,000$ measurements at each $\beta$.  From the square-lattice
results one expects the scaling form \citep{weigel:05a},
\begin{equation}
  \ln \chi(\beta) \sim A_\chi+B_\chi(\beta-\beta_c)^{-\rho},
  \label{thermal_scaling_phi4_formula}
\end{equation}
which should hold for $\beta\rightarrow\beta_c^+$ as $N_2\rightarrow\infty$ and where
logarithmic corrections have already been omitted. We find it impossible to reliably
fit all four of the parameters involved in Eq.\ (\ref{thermal_scaling_phi4_formula})
to the available data. Varying the starting values we find a multitude of local
minima of the $\chi^2$ distribution, such that virtually any result can be ``found''
for $\beta_c$ and $\rho$ in this way. Fixing one or the other of the two parameters
at the expected values $\beta_c=\ln 2$ or $\rho=1/2$, the fits become more stable.
The dependency on the range of included values of $\beta$ is found to be rather small
and for $\beta\ge 1.25$ we arrive at the fit parameters $A_\chi = -101(4662)$,
$B_\chi = 106(4662)$, $\rho = 0.02(103)$, and $Q = 0.03$, for $\beta_c$ fixed at $\ln
2$ or at the parameters $A_\chi = -86(1083)$, $B_\chi = 324(5744)$, $\beta_c =
-11(147)$, and $Q = 0.04$, with $\rho$ fixed at $1/2$. Obviously both fits are not
very useful, such that we are finally forced to fix both parameters, $\beta_c$ and
$\rho$, at their expected values to find $A_\chi = 0.91(41)$, $B_\chi = 4.20(33)$,
and $Q = 0.03$.  This fit is shown in Fig.\ \ref{phi4_thermal_scaling_fit} together
with the simulation data.  Thus, the best we can conclude about the thermal scaling
behaviour of the polarisability is that there is no obvious contradiction with the
expectations concerning the parameters $\beta_c$ and $\rho$. However, in view of the
fact that already for the regular lattice model thermal scaling fits were not at all
easily possible \citep{weigel:05a}, this finding is probably not too astonishing.

\section{Geometrical properties}
\label{sec:geom}

The annealed nature of disorder applied to the vertex model via its placement onto
dynamical $\phi^4$ random graphs induces a back-reaction of the matter variables onto
the underlying geometry and thus a possible change in the (local and global)
geometrical properties of the graphs. Since the general mechanism of matter
back-reaction onto the graphs is the tendency to minimise interfaces between
pure-phase regions of the matter variables, a strong coupling between matter and
graph variables is generically only expected if the combined system of spin model and
underlying geometry is critical. Thus, the universal graph properties such as the
graph-related critical exponents should remain at the values of pure Euclidean
quantum gravity, unless the coupled matter system has a divergent correlation length
\citep{ambjorn:93b}. As indicators for changes of the geometry of the coupled system,
we consider the co-ordination number distribution as a typical local property, as
well as the string susceptibility exponent and the Hausdorff dimension as global
geometrical features.

\subsection{The co-ordination number distribution}

\begin{figure}[tb]
  \centering
  \includegraphics[clip=true,keepaspectratio=true,width=12cm]{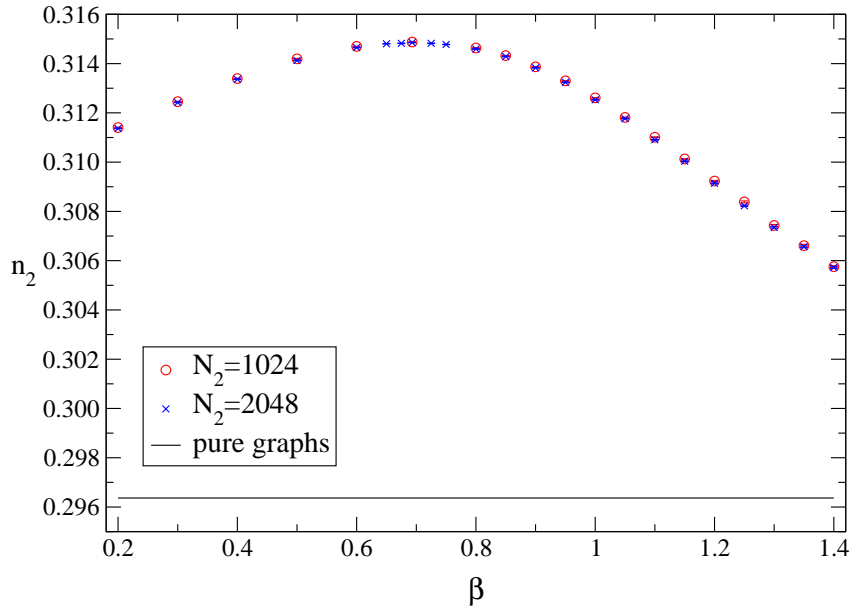}
  \caption
  {Fraction $n_2$ of faces of length two of planar $\phi^4$ ``regular'' random graphs
    with a coupled $F$ model as a function of the inverse temperature $\beta$. The
    drawn error bars are mostly covered by the size of the symbols. The solid line
    shows the value of $n_2$ for the case of pure $\phi^4$ random graphs of
    $N_2=2048$ sites.}
  \label{phi4_twoloops_peaks}
  \vspace*{0.15cm}
\end{figure}

The distribution of co-ordination numbers of the quadrangulations, which has been
extensively considered for the case of pure $\phi^4$ graphs \citep{doktor}, could be
possibly altered by the back-reaction of a coupled matter model. In particular, for
the case of the vertex model considered here, the ice rule forbids certain link-flip
update moves and thus changes the distribution $P_{N_2}(q)$ of co-ordination numbers. 
The vertex configurations forbidden by the ice rule effectively carry infinite
energy, such that they stay excluded even in the infinite-temperature limit
$\beta\rightarrow 0$. Thus, in contrast to, e.g., an Ising model a full decoupling of
graph and matter variables for high temperatures does not occur here due to the {\em
  entropic\/} instead of energetic nature of the matter-graph back reaction. 

From our numerical simulations we find that on the scale of the whole distribution
$P_{N_2}(q)$ no changes as a function of the inverse simulation temperature $\beta$
can be distinguished and the distribution looks identical to that of pure planar
$\phi^4$ graphs \citep{doktor}. However, $P_{N_2}(q)$ can be determined to high
precision, and concentrating on a single point of the distribution, e.g., $q=2$, a
clear variation with the inverse temperature $\beta$ can be resolved, cf.\ Fig.\
\ref{phi4_twoloops_peaks}. Also, in terms of the quoted statistical errors, which are
of the order of $10^{-5}$ for the measurements of $n_2\equiv P_{N_2}(2)$, the pure
graph result of $n_2=0.296\,365(32)$ for $N_2=2048$ \citep{doktor} is very far away
from the whole of the shown variation of the $F$ model case. We find a peak of $n_2$
around $\beta\approx 0.7$ with only rather small variations with the size of the
considered graph. A similar peak of the fraction $P_{N_2}(3)$ of {\em three}-faces
for different spin models coupled to dynamical {\em triangulations\/} has been
observed before, see Ref.\ \citep{baillie}.

Since a pronounced back-reaction of the matter variables onto the underlying graphs
is only expected at criticality, we interpret the location of the observed peak of
$n_2(\beta)$ as a pseudo-critical point $\beta_{n_2}$ which should scale to the
asymptotic critical coupling $\beta_c=\ln 2$. As for the thermal scaling analysis of
Sec.~\ref{sec:results1}, the precise location of the maxima can be determined from
the simulation data via reweighting.  This has been done for the data from
simulations of graphs of sizes between $N_2=256$ and $N_2=4096$ sites with time
series of lengths between $8\times 10^5$ and $4\times 10^6$ measurements. We find
only very small changes of this peak position on variation of the size of the graphs,
such that within the present statistical errors $\beta_{n_2}$ can be considered
constant.  Thus, we do not perform a finite-size scaling fit to the data of the peak
locations, but instead quote the result from the largest considered lattice as an
estimate for the asymptotic critical coupling, namely
\begin{equation}
  \beta_{n_2} = 0.6894(54),
\end{equation}
resulting from the simulations for $N_2=4096$. This is in nice agreement with the
expected value of $\beta_c=\ln 2\approx0.693$ and almost two orders of magnitude more
precise than the results found above from the scaling of the polarisability peak
locations. 

\subsection{The string susceptibility exponent}

In the grand-canonical ensemble of the dynamical polygonifications model the string
susceptibility exponent $\gamma_s$ governs the leading singularity of the partition
function for spherical graphs via $Z(\mu) \sim (\mu-\mu_0)^{2-\gamma_s}$
\citep{ambjorn:book}, where $\mu$ denotes the chemical potential accounting for the
cost of the insertion of a new vertex. Thus, a direct measurement of $\gamma_s$
requires computationally demanding simulations with a varying number of polygons or
graph vertices.  Additionally, since a shift of $\gamma_s$ due to the presence of
some matter variables can only be expected at criticality, a numerical setup for the
detection of such a change needs to tune two coupling constants, namely $\mu$ and
$\beta$, to criticality. Due to the combination of these two problems a reliable
estimation of $\gamma_s$ from grand-canonical Monte Carlo simulations has proved
difficult, see e.g.\ Ref.~\citep{ambjorn:86}.

It could be shown, however, that the string susceptibility exponent is related to the
``baby-universe'' structure of the dynamical polygonifications \citep{jain:92a}. This
observation can be turned into a method for the determination of $\gamma_s$ from
simulations at a fixed number of polygons or graph vertices (canonical ensemble)
\citep{ambjorn:93b}. The basic building blocks of this ```baby-universe'' structure
are taken as so-called ``minimal-neck baby universes'' (minBUs), which we define as
subgraphs which typically contain a ``macroscopic'' number of vertices, but are
connected to the main graph body by only four links for the case of dynamical
quadrangulations. A simple decomposition argument of the graphs into ``baby
universes'' yields the following scaling relation for the distribution $\l
n_{N_2}(B)\r$ of volumes $B$ contained in minBUs of the ensemble of pure graphs of
size $N_2$ \citep{ambjorn:93b},
\begin{equation}
  \l n_{N_2}(B)\r \sim N_2^{2-\gamma_s} [B(N_2-B)]^{\gamma_s-2},
  \label{minBU_gamma}
\end{equation}
where $B\gg 1$ and $N_2-B\gg 1$ is assumed. Also, it can be shown that the same
relation should hold for the case of $C<1$ conformal matter coupled to the
polygonifications or dual graphs with $\gamma_s$ then denoting the corresponding
dressed string susceptibility exponent \citep{jain:92a}.  For the limiting case $C=1$,
on the other hand, it is argued in Ref.\ \citep{jain:92a} that the distribution of
minBUs should acquire logarithmic corrections and look like,
\begin{equation}
  \l n_{N_2}(B)\r \sim N_2^{2-\gamma_s} [B(N_2-B)]^{\gamma_s-2}
  [\ln B\,\ln (N_2-B)]^\kappa,
  \label{minBU_gamma_c1}
\end{equation}
with $\kappa=-2$. An estimate $\bar{n}_{N_2(B)}$ for the volume distribution of
minBUs can be easily found numerically from a decomposition of the graphs into ``baby
universes''. When the minBU surgery algorithm (cf.\ Refs.\ \citep{prep,doktor}) is
applied, such an estimate can even be produced as a simple by-product of the updating
scheme. Then, an estimate for $\gamma_s$ can be found from a fit of the conjectured
functional form (\ref{minBU_gamma}) or (\ref{minBU_gamma_c1}) to the estimated
distribution $\bar{n}_{N_2(B)}$ \citep{ambjorn:93b}. In order to honour the
constraints $B\gg 1$ and $N_2-B\gg 1$ of Eqs.\ (\ref{minBU_gamma}) and
(\ref{minBU_gamma_c1}) one has to introduce cut-offs $B_\mathrm{min}$ and
$B_\mathrm{max}$, such that only data with $B_\mathrm{min}\le B\le B_\mathrm{max}$
are included in the fit. Here, the choice of the lower cut-off $B_\mathrm{min}$ is
found to be much more important for the outcome of the fit than the choice of
$B_\mathrm{max}$. We use the following recipe for the determination of the cut-offs:
as a rule of thumb, we choose $B_\mathrm{max}=N_2/8$, which has turned out to be a
good initial guess for most situations.  With $B_\mathrm{max}$ fixed, the lower
cut-off $B_\mathrm{min}$ is steadily increased from $B_\mathrm{min}\approx 0$,
monitoring the effect of those increases on the resulting fit parameters, especially
the estimated string susceptibility exponent $\gamma_s$.  Finally, with the resulting
value of $B_\mathrm{min}$ fixed, a second adaption of $B_\mathrm{max}$ is attempted,
usually changing $B_\mathrm{max}$ by factors of two or one half. Additionally,
the quality-of-fit parameter $Q$ is utilised as an indicator of whether neglected
corrections to scaling are important for the considered window of minBU volumes $B$.
As far as corrections to the leading scaling behaviour are concerned, it is
speculated in Ref.\ \citep{ambjorn:93b} that a good effective description of the
leading correction term results from the substitution $B^{\gamma_s-2}\rightarrow
B^{\gamma_s-2}[1+D_{\gamma_s}/B]$. Hence, the actual fits were performed to the
functional form
\begin{equation}
  \ln \bar{n}_{N_2}(B) = A_{\gamma_s}+(\gamma_s-2)\ln\left[B(N_2-B)\right]+
  \frac{D_{\gamma_s}}{B},
  \label{gammas_fit_form1}
\end{equation}
for $C<1$, resp.\ to the form
\begin{equation}
  \hspace*{-0.85cm}
  \ln \bar{n}_{N_2}(B) = A_{\gamma_s}+(\gamma_s-2)\ln[B(N_2-B)]
  +\kappa\ln[\ln B\ln(N_2-B)]+\frac{D_{\gamma_s}}{B},
 \label{gammas_fit_form2}
\end{equation}
for the limiting case of $C=1$. Here, the dependency on the total volume $N_2$ has
been condensed into the constant $A_{\gamma_s}$. Note that both of these fits are
linear and the number of data points is of the order of $10^3$ for the lattice sizes
we have considered, such that a fit with four independent parameters is not
unrealistic. In Eq.\ (\ref{gammas_fit_form2}) we keep $\kappa$ as a free parameter
since the value $\kappa=-2$ is only a conjecture and, additionally, further
corrections to scaling can be covered in an effective way by letting $\kappa$ vary.

\subsubsection{Results for pure $\phi^4$ graphs}

Matrix model calculations for pure, planar dynamical triangulations yield the exact
result $\gamma_s=-1/2$, cf.\ Ref.~\citep{ambjorn:book}. As a gauge for the method and
as a check for the expected universality of $\gamma_s$ with respect to the change
from triangulations to quadrangulations, we apply the described technique first to
the case of pure $\phi^4$ random graphs. We adapt the lower and upper cut-offs
$B_\mathrm{min}$ and $B_\mathrm{max}$ iteratively as described above, taking into
account that the usual error estimates of least-squares fits of
(\ref{gammas_fit_form1}) to the data could be misleading due to the apparent
correlations of the points of $\bar{n}_{N_2}(B)$ for different sizes $B$ of the
minBUs, which generically lead to an underestimation of variances. We refrain from an
additional extrapolation of the resulting estimates of $\gamma_s$ towards
$B_\mathrm{min}\rightarrow\infty$ suggested by the authors of Ref.\
\citep{ambjorn:93b} since we do not see a proper justification for a specific
extrapolation ansatz and in general find extrapolations of noisy (and here also
strongly correlated) data questionable.

\begin{table}[tb]
  \centering
  \caption
  {Parameters of fits of (\ref{gammas_fit_form1}) to the
    simulation data for the distribution $\bar{n}_{N_2}(B)$ of minBUs for {\em pure\/} $\phi^4$
    random graphs. The small values of the quality-of-fit parameter $Q$ for the two
    largest graph sizes are a side effect of the cross-correlations in $\bar{n}_{N_2}(B)$.} 
  \vspace*{0.25cm}
  \begin{tabular}{|r|r|r|r@{.}l|r@{.}l|r@{.}l|r|} \hline
    \multicolumn{1}{|c|}{$N_2$} & \multicolumn{1}{|c|}{$B_\mathrm{min}$}
    & \multicolumn{1}{|c|}{$B_\mathrm{max}$} & \multicolumn{2}{|c|}{$A_{\gamma_s}$} &
    \multicolumn{2}{|c|}{$\gamma_s$} & \multicolumn{2}{|c|}{$D_{\gamma_s}$} &
    \multicolumn{1}{|c|}{$Q$} \\ \hline \hline
     1024 & 60 & 128 & 18&36(49) & $-0$&474(40) & $-2$&9(30) & 0.79 \\
     2048 & 70 & 256 & 20&34(14) & $-0$&495(10) & $-3$&8(12) & 0.56 \\
     4096 & 70 & 512 & 22&030(90) & $-0$&4915(63) & $-3$&78(74) & 0.05 \\
     8192 & 100 & 1024 & 23&853(72) & $-0$&4977(47) & $-4$&80(87) & 0.04 \\ \hline
  \end{tabular}
  \vspace*{0.15cm}
  \label{phi4_pure_minbus_table2}
\end{table}

Statistically reliable error estimates for $\gamma_s$ are found by jackknifing over
the whole fitting procedure: first the upper and lower cut-offs in $B$ are determined
as described using the full estimate $\bar{n}_{N_2}(B)$. Then, of the order of ten
jackknife blocks are built from the time series the estimate $\bar{n}_{N_2}(B)$ is
based on and fits with the same constant cut-offs are performed for each block to
yield jackknife-block estimates of $\gamma_s$ and the other fit parameters. Table
\ref{phi4_pure_minbus_table2} summarises the final results for pure $\phi^4$ graphs
of sizes $N_2=1024$ up to $N_2=8192$, taking about $10^9\times N_2$ minBUs into
account for each graph size. Obviously, finite-size effects are relatively weak here,
and we quote as final result the value for $N_2=8192$, $\gamma_s=-0.4977(47)$, which
is perfectly compatible with $\gamma_s=-1/2$. 

\subsubsection{Results for the $F$ model case}

\begin{figure}[tb]
  \centering
  \includegraphics[clip=true,keepaspectratio=true,width=12cm]{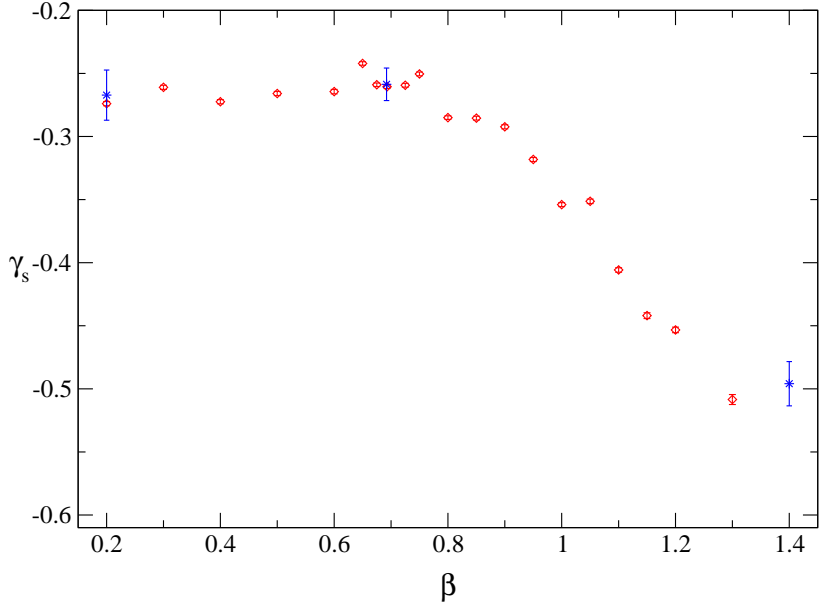}
  \caption
  {Estimates of the string susceptibility exponent $\gamma_s$ from fits of
    (\ref{gammas_fit_form1}) to the measured distribution of minBUs for graphs of
    size $N_2=2048$ coupled to the $F$ model. The displayed error bars do not
    represent the full statistical error. The true amount of statistical fluctuations
    is indicated by the three data points with larger error bars at couplings
    $\beta=0.2$, $\beta=\ln 2$, and $\beta=1.4$, where the errors have been evaluated
    by a full jackknife analysis. Note that the displayed exponent estimates in the
    high-temperature phase are {\em effective\/} exponents since there are large
    finite-size corrections (see text).}
  \label{phi4_gamma_overview}
  \vspace*{0.15cm}
\end{figure}

For the case of the $F$ model coupled to the $\phi^4$ graphs, we expect a variation
of the string susceptibility exponent $\gamma_s$ with the inverse temperature $\beta$
of the $F$ model. Since the whole high-temperature phase is critical with central
charge $C=1$, in the thermodynamic limit $\gamma_s$ should vanish for all $\beta\le
\beta_c=\ln 2$, whereas in the non-critical ordered phase the exponent should stick
to the pure quantum gravity value of $\gamma_s=-1/2$. To get an overview of the
temperature dependence of $\gamma_s$ we measured the distribution $\bar{n}_{N_2}(B)$
of minBUs over an inverse temperature range of $0.2\le\beta\le 1.4$ for graphs of
size $N_2=2048$ and performed fits of the functional form (\ref{gammas_fit_form1}) to
the data to extract $\gamma_s$ [thus first neglecting the possibility of additional
logarithmic corrections indicated in Eq.~(\ref{gammas_fit_form2})]. The resulting
estimates for $\gamma_s$ presented in Fig.\ \ref{phi4_gamma_overview} show a plateau
value of $\gamma_s\approx -0.25$ within the critical phase $\beta\le\ln 2$ and a slow
drop down to $\gamma_s\approx -0.5$ at $\beta=1.4$ in the low-temperature phase. Note
that the error bars displayed in Fig.\ \ref{phi4_gamma_overview} are those resulting
from the fit procedure itself and are thus not representing the full statistical
variation due to the above mentioned cross-correlations between the values of
$\bar{n}_{N_2}(B)$. For comparison the correct error bars as obtained from a more
elaborate jackknife analysis are shown for three selected $\beta$-values, which are
discussed in more detail below. As will be shown there, the fact that $\gamma_s$ is
found to be still considerably smaller than zero in the high-temperature phase is due
to a finite-size effect. We do not employ the corrected fit (\ref{gammas_fit_form2})
at this point, which is found to be unstable for the small graph size considered
here.

More precise estimates for $\gamma_s$ are found from a FSS study of three series of
simulations, one at the critical point $\beta_c=\ln 2$, one in the critical
high-temperature phase at $\beta=0.2$ and one deep in the ordered phase at
$\beta=1.4$. For the latter case, the exponential slowing down of the combined
link-flip and surgery dynamics of the graphs reported in Refs.~\citep{prep,doktor}
limited the maximum accessible graph size to $N_2=16\,384$, while for the simulations
at the critical point and in the high-temperature phase graphs with up to
$N_2=65\,536$ sites were considered. At $\beta=1.4$, this maximal size is anyway
sufficient since we find no finite-size drift in the estimate for $\gamma_s$ with
increasing graph sizes, all results being compatible with the conjectured value of
$\gamma_s=-1/2$. Thus, as our final estimate for $\beta=1.4$ we report the value
found for $N_2=16\,384$, $\gamma_s = -0.478(17)$. For the quoted statistical error
estimates the jackknifing procedure described above for pure dynamical $\phi^4$
graphs was used, thus taking full account of the present fluctuations.

At the critical point $\beta_c=\ln 2$ fits of the form (\ref{gammas_fit_form1})
without logarithmic corrections show considerable finite-size effects, with
$\gamma_s$ slowly increasing with the graph size.  For the largest graph size
considered, $N_2 = 65\,536$, the thus found estimate $\gamma_s=-0.2075(17)$ is still
far away from the expected result $\gamma_s=0$. Taking the logarithmic corrections
into account, however, these results can be considerably improved, with the numerical
estimates for $\gamma_s$ now being fully consistent with the theoretical prediction.
The parameters of fits of the corresponding functional form (\ref{gammas_fit_form2})
are collected in Table \ref{phi4_F_minbus_table2}. Including this correction, no
further finite-size dependence of the estimate $\gamma_s$ is visible. The occurring
values for the ``correction exponent'' $\kappa$ are not too far away from and indeed
statistically compatible with the conjectured value of $\kappa=-2$. Since for the
case of $N_2=65\,536$ only a much shorter time series than for the smaller graph
sizes was recorded, we present as our final estimate of the critical value of
$\gamma_s$ the result at $N_2=32\,768$, $\gamma_s = 0.013(70)$.

\begin{table}[tb]
  \centering
  \caption
  {Parameters of fits of the form (\ref{gammas_fit_form2}) 
    to the distribution $\bar{n}_{N_2}(B)$ of minBUs for $\phi^4$
    random graphs coupled to the $F$ model at $\beta=\beta_c=\ln 2$.} 
  \vspace*{0.25cm}
  \begin{tabular}{|r|r|r|r@{.}l|r@{.}l|r@{.}l|r@{.}l|} \hline
    \multicolumn{1}{|c|}{$N_2$} & \multicolumn{1}{|c|}{$B_\mathrm{min}$}
    & \multicolumn{1}{|c|}{$B_\mathrm{max}$} & \multicolumn{2}{|c|}{$A_{\gamma_s}$} &
    \multicolumn{2}{|c|}{$\gamma_s$} & \multicolumn{2}{|c|}{$\kappa$} & 
    \multicolumn{2}{|c|}{$D_{\gamma_s}$}
    \\ \hline \hline
    16\,384 & 100 & 2048 & 25&7(15) & $0$&05(13) & $-1$&97(89) & $-10$&9(69) \\ 
    32\,768 & 110 & 4096 & 27&08(93) & $0$&013(70) & $-1$&80(50) & $-12$&6(47) \\
    65\,536 & 120 & 4096 & 27&5(14) & $-0$&05(12) & $-1$&27(82) & $-6$&9(71) \\ \hline
  \end{tabular}
  \vspace*{0.15cm}
  \label{phi4_F_minbus_table2}
\end{table}

Finally, in the high-temperature phase at $\beta=0.2$ the simulation results behave
very similar to the critical point case. When applying fits of the form
(\ref{gammas_fit_form1}) without logarithmic corrections, considerable finite-size
effects are found, and the approach of the resulting exponent estimates $\gamma_s$ to
the expected value of $\gamma_s=0$ is very slow. On the other hand, the estimates
resulting from fits of the form (\ref{gammas_fit_form2}) to the data are compatible
with $\gamma_s=0$ for the larger of the considered graph sizes. For $N_2=32\,768$ we
find $\gamma_s = -0.041(73)$, $\kappa = -1.38(47)$, $Q = 0.05$ with cut-offs
$B_\mathrm{min}=100$ and $B_\mathrm{max}=2048$. To complete the picture, it should be
mentioned that the functional form (\ref{gammas_fit_form2}) does {\em not\/} fit the
data in the low-temperature phase at $\beta=1.4$ well and does not give estimates of
$\gamma_s$ compatible with $\gamma_s=0$, in agreement with theoretical expectations.

\subsection{The Hausdorff dimension\label{phi4_F_hausdorff_sec}}

The internal Hausdorff dimension $d_h$ of the dynamical polygonifications is one of
its most striking features.  Apart from the physical implications, its large value
causes a quite inconvenient obstacle for the numerical analysis of the model, namely
the comparable smallness of the effective linear extent of the graphs at a given
total volume $N_2$ as compared to flat lattices. As matter variables are coupled to
the dynamical graphs, the strong coupling between graph and matter variables at
criticality could lead to a change of the fractal dimension of the lattices. In a
phenomenological scaling picture, such a strong coupling of matter and geometry
should set in as soon as the correlation length of the matter system becomes
comparable to the intrinsic length scale of the graphs or polygonifications. For
conformal minimal matter, there has been quite some debate about how $d_h$ should
depend on the central charge $C$ of the coupled matter system, see, e.g., Refs.\
\citep{catterall:95a,ambjorn:95d,ambjorn:95e,ambjorn:97b,ambjorn:98b,ambjorn:98d}.
For $C=0$ the result $d_h=4$ is exact \citep{kawai}. Furthermore, the branched
polymer model \citep{ambjorn:86}, describing the $C\rightarrow\infty$ limit
\citep{david:97a}, yields $d_h=2$ (see, e.g., Ref.\ \citep{jurkiewicz:97a}).  For the
intermediate region $0\le C\le 1$ two differing conjectures have been made for $d_h$,
namely \citep{watabiki:93a}
\begin{equation}
  d_h = 2\frac{\sqrt{25-C}+\sqrt{49-C}}{\sqrt{25-C}+\sqrt{1-C}}\;
  \stackrel{C\rightarrow 1}{\longrightarrow}\; 2(1+\sqrt{2})\approx 4.83,
  \label{hausdorff_pred1}
\end{equation}
and \citep{ishibashi:94a}
\begin{equation}
  d_h = \frac{24}{\sqrt{1-C}(\sqrt{1-C}+\sqrt{25-C})}\; \stackrel{C\rightarrow
    1}{\longrightarrow}\; \infty. 
  \label{hausdorff_pred2}
\end{equation}
All numerical investigations up to now, on the other hand, are consistent with a
constant $d_h=4$ for $0\le C\le 1$
\citep{ambjorn:95d,ambjorn:95e,ambjorn:97a,ambjorn:98d}. Naturally, the limiting case
$C=1$ considered here is of special interest for the investigation of the transition
to the branched polymer regime $C\gg1$. Numerically, it has proved exceptionally
difficult to extract the Hausdorff dimensions from the statistics of the practically
accessible graph sizes \citep{billoire:86a,agishtein:91b,ambjorn:95e}. Only more
recently, the development and application of suitable FSS techniques allowed for a
more successful and precise determination of $d_h$
\citep{catterall:95a,ambjorn:97b,ambjorn:98d}.

\subsubsection{Scaling and the two-point function}

The fractal structure of the polygonifications is encoded in their geometrical
two-point function. Here, different definitions are possible. While in
Eq.~(\ref{msqe_def_sim}) a definition in terms of the vertices of the graphs has been
used, here, instead, the number of vertices of the quadrangulation is counted.  Thus,
we define the geometrical two-point function $G_{11}^{N_2}(r)$ as the average number
of vertices of the polygonifications at a distance $r$ from a marked vertex, where
``distance'' denotes the unique minimal number of links one has to traverse to
connect both vertices. Since the intrinsic length of the model scales as
$N_2^{1/d_h}$ by definition of the internal Hausdorff dimension $d_h$, from the usual
FSS arguments one can make the following scaling ansatz (see, e.g., Ref.\
\citep{catterall:95a}),
\begin{equation}
  G_{11}^{N_2}(r) \sim N_2^{\alpha}\,F(r/N_2^{1/d_h}), 
  \label{twopoint_scaling_ansatz}
\end{equation}
i.e., $G_{11}^{N_2}(r)$ is a generalised homogeneous function and one can define a
scaling function $F(x)$ of the single scaling variable $x=r/N_2^{1/d_h}$ and a
critical exponent $\alpha$. Due to the obvious constraint $N_2 = \sum_{r}
G_{11}^{N_2}(r)$, the exponent $\alpha$ is not independent, but given by
$\alpha=1-1/d_h$. It turns out that for practical purposes the scaling variable has
to be {\em shifted\/} to yield reliable results, see, e.g., Refs.\
\citep{ambjorn:97b,ambjorn:98b,ambjorn:99a}.  The necessity of such a shift can be
most easily seen by a phenomenological scaling discussion of the mean extent defined
by
\begin{equation}
  \l r\r_{N_2} = \frac{1}{N_2}\sum_r r\,G_{11}^{N_2}(r)\sim F_0N_2^{1/d_h},
  \label{mean_extent_scaling}
\end{equation}
with $F_0 = \sum_r F(r)$. On general grounds, one expects the presence of analytical
scaling corrections,
\begin{equation}
  \frac{\l r\r_{N_2}}{N_2^{1/d_h}} \sim
  F_0-\frac{a}{N_2^{1/d_h}}-\frac{b}{N_2^{2/d_h}}+\cdots. 
\end{equation}
Combining the terms proportional to $1/N_2^{1/d_h}$ on both sides, the mean extent is
found to be $\l r+a\r_{N_2} \sim F_0N_2^{1/d_h}+O(N_2^{-1/d_h})$.  Thus, to
incorporate first-order corrections to scaling, the ansatz
(\ref{twopoint_scaling_ansatz}) is replaced by
\begin{equation}
  G_{11}^{N_2}(r) \sim N_2^{\alpha}\,F[(r+a)/N_2^{1/d_h}],
  \label{twopoint_scaling_ansatz_shifted}
\end{equation}
i.e., the scaling variable is now defined to be $x=(r+a)/N_2^{1/d_h}$. 

\subsubsection{Scaling of the maxima}

The two-point function $G_{11}^{N_2}(r)$ exhibits a peak at intermediate distances
and declines exponentially as $r\rightarrow\infty$, cf.\ Fig.\
\ref{phi4_twopoint_collapse} below. From the scaling ansatz
(\ref{twopoint_scaling_ansatz_shifted}) one infers the following leading scaling
behaviour of the position and height of the maxima,
\begin{equation}
  \begin{array}{rcl}
    r_\mathrm{max}+a & = & A_r N_2^{1/d_h}, \\
    G_{11}^{N_2}(r_\mathrm{max}) & = & A_n N_2^{1-1/d_h}+B_n. 
  \end{array}
  \label{hausdorff_peaks_fitform}
\end{equation}
Since the location and height of these maxima can be determined numerically from
simulation data, these relations can be used to estimate the intrinsic Hausdorff
dimension $d_h$. A technical difficulty is given by the fact that $r$ can only take
on integer values for the discrete graphs considered. This problem is circumvented by
a smoothing out of the vicinity of the maximum by a fit of a low-order polynomial to
$G_{11}^{N_2}(r)$ around its maximum. For practical purposes, we find a fourth-order
polynomial sufficient for this fit. Reliable error estimates are found by jackknifing
over this whole fitting procedure, where the individual statistical errors of the
data points included in the fits are taken to be equal. Thus, one arrives at
estimates for the peak locations $r_\mathrm{max}$ and heights
$G_{11}^{N_2}(r_\mathrm{max})$ as a function of the graph size $N_2$, to which then
the functional forms of Eq.\ (\ref{hausdorff_peaks_fitform}) are fitted. The effect
of neglected FSS corrections is accounted for by successively dropping data points
from the small-$N_2$ side. For simulations of pure $\phi^4$ random graphs, in this
way we find the value of $d_h$ to steadily increase on omitting more and more points.
For the range $N_2=4096$ up to $N_2=32\,768$ we thus arrive at the estimates $d_h =
3.803(28)$, $Q=0.22$ from the scaling of the peak locations and $d_h = 3.814(63)$,
$Q=0.44$ from the peak heights. Both estimates are still noticeably away from the
asymptotic values $d_h=4$, owing to the neglect of higher-order correction terms
\citep{catterall:95a}. We note that introducing the shift parameter $a$ already
largely improved the estimates, since fixing $a=0$ we arrive at $d_h=3.4313(20)$ from
the peak locations. Further improvement is gained from the inclusion of the
next-order correction term for the scaling of the peak locations,
\begin{equation}
    r_\mathrm{max}+a = A_r N_2^{1/d_h}+B_r N_2^{-1/d_h},
  \label{hausdorff_peaks_fitform2}
\end{equation}
which yields an estimate of $d_h = 3.964(42)$, $Q = 0.24$ for the range
$N_2=512,\ldots,32\,768$, in perfect agreement with $d_h=4$. 

\begin{figure}[tb]
  \centering
  \includegraphics[clip=true,keepaspectratio=true,width=12cm]{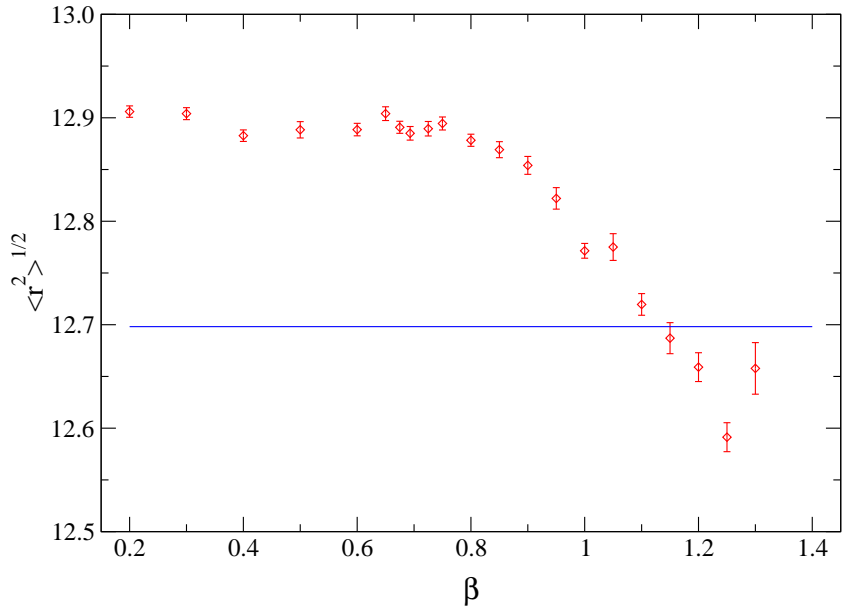}
  \caption
  {Root mean square extent $\l r^2\r^{1/2}$ of regular $\phi^4$ random graphs with
    $N_2=2048$ sites coupled to the $F$ model. The horizontal line indicates the root
    mean square extent of pure $\phi^4$ random graphs of the same size.}
  \label{phi4_msqe_overview}
  \vspace*{0.15cm}
\end{figure}

For random $\phi^4$ graphs coupled to the $F$ model, we find a small dependence of
the root mean square extent on the inverse temperature $\beta$ of the coupled $F$
model and also a slight shift of $\l r^2\r^{1/2}$ as compared to the case of pure
$\phi^4$ random graphs, cf.\ Fig.\ \ref{phi4_msqe_overview}. Thus, one might expect
the Hausdorff dimension $d_h$ to be temperature dependent, too. We performed
simulations for three inverse temperatures, namely $\beta=0.2$, $\beta=\ln 2$ and
$\beta=1.4$, covering the cases of interest.  The results for $d_h$ from fits of the
functional form (\ref{hausdorff_peaks_fitform}) to the data are found to steadily
increase on omitting more and more points from the small-$N_2$ side. In agreement
with the case of pure $\phi^4$ graphs, the final estimates for $d_h$ are found to be
significantly smaller than $d_h=4$ for all three inverse temperatures due to the
presence of higher-order corrections to scaling. For the peak heights, this analysis
yields the estimates $d_h = 3.446(68)$ for $\beta = 0.2$, $d_h = 3.426(92)$ for
$\beta=\ln 2$ and $d_h = 3.94(23)$ for $\beta = 1.4$, where the rather different
result for $\beta = 1.4$ again indicates the presence of competing local minima in
the $\chi^2$ distribution. The found higher-order scaling corrections are resolved by
using the fit ansatz (\ref{hausdorff_peaks_fitform2}) for the peak locations. Here,
we do not find a significant sensitivity of the parameter estimates on the cut-off
$N_{2,\mathrm{min}}$ and for all three inverse temperatures the resulting values for
$d_h$ are in agreement with the pure gravity value $d_h=4$, cf.\ the fit data collected
in Table \ref{phi4_F_hausdorff_table3}.

\begin{table}[tb]
  \centering
  \caption
  {Parameters of fits of the form (\ref{hausdorff_peaks_fitform2})
    to the peak locations of the two-point functions of the random graph $F$ model. The
    maximum graph size was $N_2=65\,536$ for $\beta=0.2$ and $\beta=\ln2$
    and $N_2=32\,768$ for $\beta=1.4$.} 
  \vspace*{0.25cm}
  \begin{tabular}{|c|r|r@{.}l|r@{.}l|r@{.}l|r@{.}l|r@{.}l|} \hline
    \multicolumn{1}{|c|}{$\beta$} & \multicolumn{1}{|c|}{$N_{2,\mathrm{min}}$} &
    \multicolumn{2}{|c|}{$A_r$} & \multicolumn{2}{|c|}{$B_r$} & 
    \multicolumn{2}{|c|}{$a$} & \multicolumn{2}{|c|}{$d_h$} &
    \multicolumn{2}{|c|}{$Q$} \\ \hline \hline
    0.2 & 2048 & 2&34(43) & 9&2(39) & 5&5(21) & 4&13(20) & 0&86 \\
    $\ln 2$ & 2048 & 1&96(39) & 5&6(43) & 3&7(20) & 3&93(21) & 0&11 \\
    1.4 & 1024 & 2&18(49) & 6&0(37) & 4&4(22) & 4&07(26) & 0&44  \\ \hline
  \end{tabular}
  \vspace*{0.15cm}
  \label{phi4_F_hausdorff_table3}
\end{table}

\subsubsection{Scaling of the mean extent}

As an alternative to the scaling of the maxima of the two-point function, one can
also consider the behaviour of {\em mean\/} properties of the distribution
$G_{11}^{N_2}(r)$, especially the scaling of the mean extent
(\ref{mean_extent_scaling}). Taking the next sub-leading analytic correction term
into account, we make the scaling ansatz
\begin{equation}
  \l r+a\r_{N_2} = A_{\l r\r}N_2^{1/d_h}+B_{\l r\r}N_2^{-1/d_h}. 
  \label{mean_extent_scaling2}
\end{equation}
We again consider the case of pure $\phi^4$ graphs first. When fixing $B_{\l r\r}=0$
and adapting the lower cut-off $N_{2,\mathrm{min}}$, the resulting values of $d_h$
are significantly too small in terms of the statistical errors with an obvious
tendency to increase as more and more of the points from the small-$N_2$ side are
omitted. On the other hand, including the correction term of Eq.\
(\ref{mean_extent_scaling2}) largely reduces the dependency on the cut-off
$N_{2,\mathrm{min}}$. For $N_{2,\mathrm{min}}=256$ we find $d_h = 3.90(15)$,
$Q=0.01$, in nice agreement with $d_h = 4$. Here, the fits become very unstable as
less points are included; this explains the use of the cut-off
$N_{2,\mathrm{min}}=256$, although the quality-of-fit is rather poor. 

The authors of Ref.\ \citep{ambjorn:97b} have proposed a different and less
conventional method to extract $a$ and $d_h$ from data of the mean extent, which they
claim to be especially well suited for obtaining high-precision results. They
consider the combination $R_{a,N_2}(d_h) \equiv \l r+a\r_{N_2}N_2^{-1/d_h}$, and
evaluate it for a series of simulations for different graph sizes $N_2$. Then, for a
given $a$ and for each pair $(N_2^i,N_2^j)$ they define $d_h^{ij}(a)$ such that
$R_{a,N_2^i}(d_h^{ij})=R_{a,N_2^j}(d_h^{ij})$, i.e.,
\begin{equation}
  d_h^{ij}(a) = \frac{\ln N_2^i -\ln N_2^j}{\ln
    (\l r\r_{N_2^i}+a)-\ln (\l r\r_{N_2^j}+a)}. 
\end{equation}
By a binning technique, an error estimate $\sigma(d_h^{ij})$ is evaluated and the
estimates $d_h^{ij}(a)$ are averaged over all pairs $(N_2^i,N_2^j)$ of volumes,
$\bar{d}_h(a) = \sum_{i<j}d_h^{ij}(a)/{\mathcal N}$, where ${\mathcal N}$ denotes the
number of pairs $(N_2^i,N_2^j)$. Then, the optimal choice $a_\mathrm{opt}$ of the
shift is found by minimising
\begin{equation}
  \chi^2(a) = \sum_{i<j}\frac{[d_h^{ij}(a)-\bar{d}_h(a)]^2}{\sigma^2[d_h^{ij}(a)]},
\end{equation}
being accompanied by an optimal estimate $\bar{d}_h(a_\mathrm{opt})$. The authors of
Ref.\ \citep{ambjorn:97b} suggest to estimate the statistical error of this final
estimate by considering the variation of $(a,\bar{d}_h)$ in an interval of $a$ around
$a_\mathrm{opt}$ defined by $\chi^2(a)<\min[1,2\chi^2(a_\mathrm{opt})]$. We
implemented this procedure to compare with the results of the fits to Eq.\
(\ref{mean_extent_scaling2}) with $B_{\l r\r}=0$ for the case of pure $\phi^4$
graphs. We find the {\em ad hoc\/} assumption for the estimation of the errors of
$(a,\bar{d}_h)$ not adequate. Instead, we apply a second-order jackknifing technique
(cf.\ Ref.~\citep{doktor}) to be able to give error estimates for $d_h^{ij}(a)$ as
well as the final estimate $(a,\bar{d}_h)$ which are found to be largely differing
from those resulting from the rule $\chi^2(a)<\min[1,2\chi^2(a_\mathrm{opt})]$,
ranging from four times smaller to ten times larger error estimates. The estimates of
$d_h$ itself are found to be indeed slightly increased as compared to the fit method
(which yielded estimates clearly smaller than $d_h = 4$). This, however, can be
traced back to the fact that the individual estimates $d_h^{ij}(a)$ all receive the
same weight in the average $\bar{d}_h(a)$ above, irrespective of their precision,
giving an extra weight to the results for larger graphs, which cannot be justified on
statistical grounds. If, instead, we use a variance-weighted average
\begin{equation}
  \bar{d}_h(a) = \frac{\sum_{i<j}d_h^{ij}(a)/\sigma^2[d_h^{ij}(a)]}
  {\sum_{i<j}1/\sigma^2[d_h^{ij}(a)]},
  \label{d_h_average_amb2}
\end{equation}
the resulting estimates for $d_h$ and $a$ are statistically equivalent to those found
from the fits to (\ref{mean_extent_scaling2}). For a cut-off $N_{2,\mathrm{min}} =
2048$, for instance, we find $\bar{d}_h = 3.97(12)$ compared to $d_h = 3.99(12)$ from
a simple fit of the form (\ref{mean_extent_scaling2}) with $B_{\l r\r}=0$. Thus, we
do not find any special benefits of this computationally rather demanding method as
compared to a plain fit to (\ref{mean_extent_scaling2}) with $B_{\l r\r}=0$ and hence
do not present further detailed results for this method.

For the case of the $F$ model coupled to the $\phi^4$ random graphs we proceeded as
before, again using simulation data for $\beta=0.2$, $\beta=\ln 2$ and $\beta=1.4$.
The results from fits of the mean extent $\l r\r_{N_2}$ to the form
(\ref{mean_extent_scaling2}) with $B_{\l r\r}=0$ show very much the same behaviour as
the results from the scaling of the maxima of the two-point function, with estimates
of $d_h$ clearly below $d_h=4$ and slowly increasing as more and more points from the
small-$N_2$ side are omitted from the fits.  The outcomes of the method of Ref.\
\citep{ambjorn:97b} described above, with the average (\ref{d_h_average_amb2}) and
the $\chi^2(a)$ rule replaced by a jackknife error estimate, are again very close to
the fit results. Including the correction term of (\ref{mean_extent_scaling2}), i.e.,
relaxing the constraint $B_{\l r\r}=0$, on the other hand, yields estimates
consistent with $d_h=4$ for $\beta=0.2$ and $\beta=1.4$, however with rather large
statistical errors, cf.\ the parameters collected in Table
\ref{phi4_F_hausdorff_table5}. Note that, as mentioned before, the results for
$\beta=1.4$ are in general less precise than those for the other two inverse
temperatures, which is due to the exponential slowing down of the combined link-flip
and surgery dynamics in the low-temperature phase, cf.\ Ref.~\citep{prep}.  The fit
for $\beta=\ln 2$ settles down at a completely different minimum of the $\chi^2$
distribution, yielding an almost unchanged $d_h$ compared to the outcome of the
corresponding fit without correction term. This underlines the fact that the
complexity of the chosen fit is at least at the verge of being too high for the
available data. Nevertheless, combining the data for $d_h$ from the presented methods
and including the comparison to the pure gravity case, we find no reason to assume
that $d_h$ differs from $d_h=4$ for the case of the $F$ model coupled to $\phi^4$
random graphs. At any rate, the values $d_h\approx 4.83$ and $d_h=\infty$ resulting
from the analytical conjectures Eqs.~(\ref{hausdorff_pred1}) and
(\ref{hausdorff_pred2}) for $C=1$, respectively, are clearly incompatible with the
results found here.

\begin{table}[tb]
  \centering
  \caption
  {Parameters of fits of the form (\ref{mean_extent_scaling2}) including the
    correction term to the mean extent of dynamical $\phi^4$ graphs coupled to the $F$
    model at different inverse temperatures $\beta$.} 
  \vspace*{0.25cm}
  \begin{tabular}{|c|r|r@{.}l|r@{.}l|r@{.}l|r@{.}l|r@{.}l|} \hline
    \multicolumn{1}{|c|}{$\beta$} & \multicolumn{1}{|c|}{$N_{2,\mathrm{min}}$} &
    \multicolumn{2}{|c|}{$A_{\l r\r}$} & \multicolumn{2}{|c|}{$B_{\l r\r}$} &  
    \multicolumn{2}{|c|}{$a$} & \multicolumn{2}{|c|}{$d_h$} &
    \multicolumn{2}{|c|}{$Q$} \\ \hline \hline
    0.2 & 512 & 2&58(48) & 11&4(33) & 7&0(22) & 4&08(21) & 0&10 \\
    $\ln 2$ & 512 & 1&37(22) & 0&4(29) & 1&1(12) & 3&45(14) & 0&41 \\
    1.4 & 512 & 2&6(10) & 9&1(58) & 6&2(42) & 4&15(47) & 0&29  \\ \hline
  \end{tabular}
  \vspace*{0.15cm}
  \label{phi4_F_hausdorff_table5}
\end{table}

\begin{figure}[tb]
  \centering
  \includegraphics[clip=true,keepaspectratio=true,width=12cm]{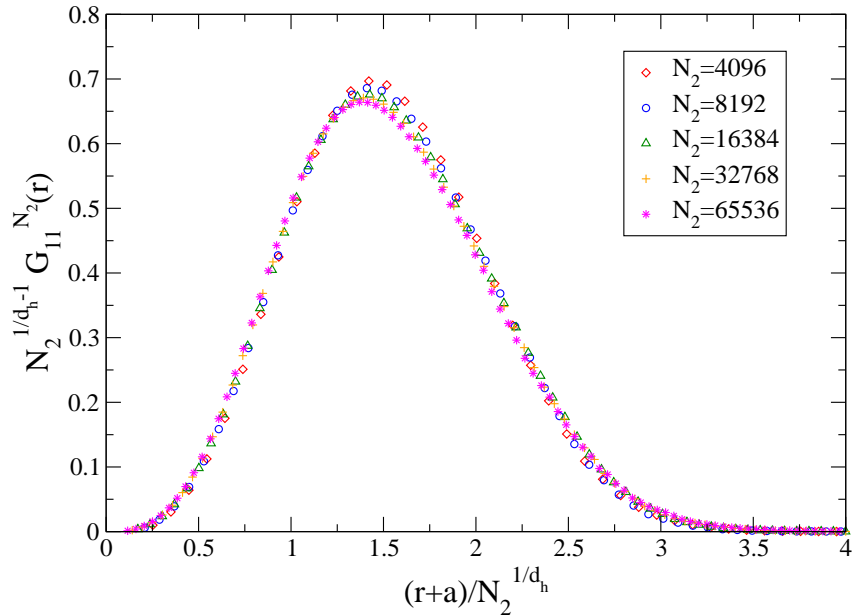}
  \caption
  {Scaling collapse of the two-point functions $G_{11}^{N_2}(r)$ of $\phi^4$ graphs
    coupled to the $F$ model at $\beta=0.2$, re-scaled according to Eq.\ 
    (\ref{twopoint_scaling_ansatz_shifted}) with $d_h=3.57$ and $a=1.60$.} 
  \label{phi4_twopoint_collapse}
  \vspace*{0.15cm}
\end{figure}

Finally, we note that the parameters $a$ and $d_h$ determined from the fits discussed
above lead to a nice scaling collapse of the two-point functions $G_{11}^{N_2}(r)$
when re-scaled according to the scaling ansatz of Eq.\
(\ref{twopoint_scaling_ansatz_shifted}). Figure \ref{phi4_twopoint_collapse} shows
this collapse of distributions for the case of $\beta=0.2$ and the choice of
parameters found from a fit to the form (\ref{mean_extent_scaling2}) with $B_{\l
  r\r}=0$, i.e., $d_h=3.57(12)$ and $a=1.60(74)$. The visible deviations around the
peaks of the distributions indicate the presence of higher-order corrections not
incorporated into the scaling ansatz (\ref{twopoint_scaling_ansatz_shifted}). 

\section{Conclusions}
\label{sec:concl}

The six-vertex $F$ model represents an example of the limiting case of a critical
theory at the ``barrier'' of central charge $C=1$, where the Liouville approach to
Euclidean quantum gravity in two dimensions breaks down \citep{kpzddk} and the
ensemble of planar random graphs coupled to such matter is at the verge of a
presumable collapse towards a phase of minimally connected, tree-like surfaces termed
``branched polymers'' \citep{david:97a}. At the same time, the family of ice-type
vertex models of statistical mechanics includes as sub-classes a variety of
well-known lattice spin models and combinatorial counting problems and hence the
analysis of its coupling to two-dimensional Euclidean quantum gravity is that of a
prototype model of statistical mechanics subject to annealed, correlated connectivity
disorder from random graphs (see also Ref.~\citep{wj:04a}). 

For studying the effect of a coupled six-vertex $F$ model, we generalised the
well-established methods of simulating dynamical triangulations to the case of planar
quadrangulations and the dual ``fat'' $\phi^4$ random graphs; the details of this
simulational machinery will be presented in a forthcoming publication \citep{prep}.
We have analysed the critical and off-critical behaviour of this model using a series
of extensive Monte Carlo simulations and subsequent finite-size and thermal scaling
analyses. On the square lattice, this model undergoes an infinite-order phase
transition of the Kosterlitz-Thouless type to an anti-ferroelectric phase of
staggered order. Expecting similar ordering behaviour to occur for the model on a
random quadrangulation, we generalised the corresponding staggered polarisation (the
order parameter) to the random graph case by a duality transformation of the vertex
model.

The scaling analysis of the simulation data is hampered by the presence of
extraordinarily strong corrections, which can be traced back to the combined effect
of the comparable smallness of the effective linear extents of the considered
(two-dimensional) lattices due to their large fractal dimension close to four and the
presence of logarithmic corrections generically expected for a $C=1$ critical point.
Additionally, the form of the critical singularities for a Kosterlitz-Thouless phase
transition severely limits the effectivity of the usual finite-size scaling
techniques. General symmetry considerations imply that the $C=1$ critical point of
the random-graph model should occur at the coupling $\beta_c=\ln 2$, which is quite
remarkably identical to the critical coupling of the square-lattice model.
Additionally, this is in agreement with a matrix model treatment of the system
\citep{zinn,zinn-justin:03a}. Due to the aforementioned strength of scaling
corrections, a precise determination of the critical coupling from the scaling of the
polarisability alone is found to be hard. A comparison of the peak positions
re-scaled according to the mean linear extents of the lattices between the random
graph and square-lattice models \citep{weigel:05a}, however, shows that the
finite-size scaling approaches of both models are indeed very similar, but with
larger correction amplitudes for the random graph model. Subsequently, however, a
precise and consistent estimate of the transition point could be extracted from the
scaling of the co-ordination number distribution of the graphs. A cursory comparison
of the scaling behaviour of the model for different ensembles regarding the inclusion
of singular contributions in the graphs reveals that corrections to scaling {\em
  increase\/} as more and more singular contributions are included. This contrasts
with the findings for pure gravity and Potts models coupled to the polygonifications
model \citep{ambjorn:95a,doktor}. As far as the critical exponents related to the
order parameter are concerned, a finite-size scaling analysis of the spontaneous
polarisation and the polarisability at the asymptotic critical coupling yields
critical exponents in good agreement with the predictions from the KPZ/DDK formula.
An attempted {\em thermal\/} scaling analysis of the polarisability around its peak
remains inconclusive due to the extraordinary magnitude of finite-size corrections.
As a curiosity, we report the finding of a critical internal energy of the model,
$U(\beta_c) = 1/3$, which is identical between the square-lattice and random graph
cases.

Several aspects of the back-reaction of the matter variables onto the properties of
the $\phi^4$ random graphs are analysed as a function of temperature. The
distribution of co-ordination numbers of the quadrangulations can be determined very
accurately. The fraction of quadrangulation sites of co-ordination number two is
found to be peaked around the asymptotic critical coupling, thus defining a
pseudo-critical point which determines the infinite-volume critical coupling very
accurately. A scaling analysis of the distribution of ``baby universes'' of the
graphs in the spirit of Refs.\ \citep{jain:92a,ambjorn:93b} allows to extract the
string susceptibility exponent $\gamma_s$ of the model. It is found to coincide with
the value $\gamma_s=0$ expected for a $C=1$ theory throughout the critical
high-temperature phase. The pure-gravity value $\gamma_s=-1/2$ is found in the
non-critical low-temperature phase. Exploiting finite-size scaling relations, we
finally analyse the geometrical two-point function of the graphs and extract the
fractal Hausdorff dimension. We find it to be consistent with the pure gravity value
$d_h=4$ for all temperatures of the coupled vertex model.  The analogous analyses for
the case of pure $\phi^4$ random graphs convincingly demonstrate the universality of
these graph-related critical exponents with respect to a change from triangulations
to quadrangulations.

In summary, despite of the presence of scaling corrections of extraordinary size, a
careful analysis of our simulation data allows for an independent confirmation of the
location of the critical point and the behaviour of the string susceptibility
exponent predicted by the matrix model treatment \citep{zinn,zinn-justin:03a}. In
addition, the behaviour and critical exponents related to the order parameter,
energy-related observables, as well as further geometrical properties such as the
Hausdorff dimension can be reliably determined. An even richer behaviour can be
expected for the 8-vertex model coupled to dynamical quadrangulations, such that its
analysis by a series of simulations similar to the one presented here would be a
promising future enterprise. 

\section*{Acknowledgements}

This work was partially supported by the EC research network HPRN-CT-1999-00161
``Discrete Random Geometries: from solid state physics to quantum gravity'' and by
the German-Israel-Foundation (GIF) under contract No.\ I-653-181.14/1999. M.W.\ 
acknowledges support by the DFG through the Graduiertenkolleg ``Quantenfeldtheorie''. 


\end{document}